\documentclass[11pt,tbtags,fleq,reqno,sumlimits,intlimits]{amsart}

\usepackage{pstricks}
\usepackage{graphicx}

\usepackage{mathrsfs}
\usepackage{latexsym}
\usepackage{amsthm,amsopn,amsfonts,amssymb,epsfig}
\numberwithin{equation}{section} \makeatletter
\renewcommand{\subsubsection}{\@startsect{subsubsection} {3} {0mm} {\baselineskip} {-0.5\baselineskip}
{\normalfont\normalsize\bfseries}} \makeatother

\newtheorem{theorem}{Theorem}
\newtheorem{lemma}[theorem]{Lemma}
\newtheorem{proposition}[theorem]{Proposition}

\newtheorem{corollary}[theorem]{Corollary}

\theoremstyle{remark}

\newtheorem*{acknow}{Acknowledgments}

\newcommand{\LLt}{\ensuremath{\widetilde{\mathcal{L}}^\alpha_\mathbf{m}}}
\newcommand{\LL}{\ensuremath{{\mathcal{L}}^\alpha_\mathbf{m}}}
\newcommand{\HHt}{\ensuremath{\widetilde{\mathcal{H}}_\mathbf{m}}}
\newcommand{\HH}{\ensuremath{{\mathcal{H}}_\mathbf{m}}}
\newcommand{\mm}{\ensuremath{\mathbf{m}}}
\newcommand{\bb}{\ensuremath{\mathbf{b}}}
\newcommand{\aaa}{\ensuremath{\mathbf{a}}}
\newcommand{\MM}{\ensuremath{|\mathbf{m}|}}
\newcommand{\order}[1]{\ensuremath{\mathrm{O}\left(#1\right)}}
\newcommand{\KKL}{\ensuremath{{\mathcal{K}}^\alpha_\mathbf{m}}}
\newcommand{\KKH}{\ensuremath{{\mathcal{K}}_\mathbf{m}}}
\newcommand{\bKKL}{\ensuremath{\bar{{\mathcal{K}}}^\alpha_\mathbf{m}}}

\newcommand{\Ait}[1]{\ensuremath{\widetilde{\mathcal{A}\dot{\imath\,}}\!_{#1}}}
\newcommand{\Ai}[1]{\ensuremath{\mathcal{A}\dot{{\imath\,}}\!_{#1}}}

\begin{document}

\title{Asymptotic correlations for Gaussian and Wishart matrices with
external source}

\author{Patrick  Desrosiers} 
\address{Department of Mathematics and Statistics, University of
Melbourne, Parkville, Victoria 3010, Australia.}
\email{P.Desrosiers@ms.unimelb.edu.au}

\author{Peter J. Forrester}
\address{Department of Mathematics and Statistics, University of
Melbourne, Parkville, Victoria 3010, Australia.}
\email{P.Forrester@ms.unimelb.edu.au}

\date{July 2006}


\maketitle

\begin{abstract}

We consider  ensembles of Gaussian (Hermite) and Wishart (Laguerre)
$N\times N$ hermitian matrices.  We study the effect of finite rank
perturbations of these ensembles by a source term.  The rank $r$ of
the perturbation corresponds to the number of non-null eigenvalues
of the source matrix.  In the perturbed ensembles, the correlation
functions can be written in terms of kernels.  We show that for all
$N$, the difference between the perturbed and the unperturbed
kernels is a degenerate kernel of size $r$ which depends on multiple
Hermite or Laguerre functions.  We also compute asymptotic formulas
for the multiple Laguerre functions kernels in terms multiple Bessel
(resp.\ Airy) functions.  This leads to the large $N$ limiting
kernels at the hard (resp.\ soft) edge of the spectrum of the
perturbed Laguerre ensemble.  Similar results are obtained in the
Hermite case.

\end{abstract}

\tableofcontents

\section{Introduction}
We  study two specific ensembles of $N\times N$ random hermitian
matrices $\mathbf{X}$ in which the probability distribution function
(p.d.f.) takes the general form
\begin{equation}\label{MatPDF}P(\mathbf{X}|\mathbf{A})\propto  e^{-\mathrm{Tr}
V(\mathbf{X})-\mathrm{Tr}\mathbf{A}\mathbf{X}},
\end{equation}
where $\mathbf{A}$ is a given $N\times N$ hermitian matrices and
$V(\mathbf{X})$ is a matrix-valued function of $\mathbf{X}$. These
matrix ensembles have been considered  by many authors, notably
Br\'ezin and Hikami \cite{Brezin},  Guhr and Wettig \cite{Guhr,GuhrWettig}, Zinn-Justin \cite{Zinn}; and more
recently,  Baik, Ben~Arous and P\'ech\'e \cite{Baik2004,Baik2005,
Peche}, El Karoui \cite{ElKaroui}, Bleher and
Kuijlaars\cite{Bleher, BleherI}, Imamura and  Sasamoto \cite{ImaSasa}.

Using the celebrated Harish-Chandra
\cite{Harish} (or Itzykson-Zuber \cite{Itzykson}) formula,
\begin{equation}\label{IZ}
\int_{\mathbf U \in U(N)} e^{{\rm Tr}(\mathbf U \mathbf A \mathbf
U^\dagger \mathbf B)} (d \mathbf U) =
\prod_{j=1}^N \Gamma(j) \frac{\det [ e^{a_i b_j} ]_{i,j=1,\dots,N}
}{\det [ a_i^{j-1}]_{i,j=1,\dots,N} \det [b_i^{j-1}]_{i,j=1,\dots,N}
}
\end{equation}
where $(d\mathbf{U})$ stands for the normalized Haar measure, one readily shows
that the eigenvalue joint p.d.f.~is
\begin{equation}\label{eigenpdf} P_N(
\mathbf{x}|\mathbf{a})=\frac{1}{Z_N}\det\left[e^{-a_j
x_k}\right]_{j,k=1}^N\prod_{i=1}^Ne^{-V(x_i)}\prod_{1\leq j<k\leq
N}\frac{x_k-x_j}{a_k-a_j},
\end{equation}
where $\mathbf{x}=(x_1,\ldots,x_N)$ denotes the random
eigenvalues of $\mathbf{X}$ while $\mathbf{a}=(a_1,\ldots,a_N)$ denotes
the fixed
eigenvalues of $\mathbf{A}$.  The quantity $Z_N$ is the normalization.
We suppose that $x_1\geq \ldots\geq x_N$
and that $x_j\in J\subseteq\mathbb{R}$.  An essential feature of
(\ref{eigenpdf}) is that it is of the form
\begin{equation}\label{DP} P_N(
\mathbf{x}|\mathbf{a})=\frac{1}{Z_N}\prod_{1\leq j<k\leq
N}\frac{1}{{a_k-a_j}}\det\left[\xi_j(x_k)\right]_{j,k=1}^N
\det\left[\eta_j(x_k)\right]_{j,k=1}^N
\end{equation}
(set $\xi_j(x)=x^{j-1}e^{-a_j x-V(x)}$, $\eta_j(x)=x^{j-1}$,
and make use of the
Vandermonde determinant evaluation in the latter).
We are thus dealing with determinantal point processes, which
can be studied as biorthogonal
ensembles in the sense of Borodin  \cite{Borodin}.

In this article, we consider the large $N$ behavior of the correlations
in two biorthogonal ensembles having an eigenvalue p.d.f.~of the form
\eqref{eigenpdf},
\begin{equation}\label{21}
\begin{array}{llll}
1)& V(x)=x^2 &J=(-\infty,\infty) & \mbox{Gaussian (Hermite) Unitary
Ensemble} ;\\
2) & V(x)=x-\alpha\ln(x) &J=(0,\infty) & \mbox{Wishart (Laguerre)
Unitary
Ensemble}.\\
\end{array}
\end{equation}
To simplify subsequent integral representations, $\alpha$ will be taken
to be a non-negative integer; however we expect our final formulas,
appropriately interpreted,
to be valid for general real $\alpha > -1$.

Both the choices in (\ref{21}) can be realized by matrices with
Gaussian entries.
Consider case 1). Let $\mathbf A$ be an $N \times N$ Hermitian matrix
with distinct eigenvalues $a_1,\dots,a_N$. Let $\mathbf Y$ be an
element of the GUE, and thus be an $N \times N$ complex Hermitian
matrix with probability density function proportional to
$e^{-{\rm Tr} \, \mathbf Y^2}$. With $t$ a positive scalar, it is a
well known consequence of (\ref{IZ}) (see e.g.~\cite{FR2}) that the eigenvalue p.d.f.~of
\begin{equation}\label{18.1}
- \mathbf A + \sqrt{t} \mathbf Y
\end{equation}
is equal to
\begin{equation}\label{S1}
\frac{(-1)^{N(N-1)/2} }{ N! (2 \pi t)^{N/2} } \det [ e^{-(\lambda_i
+ a_j)^2/2t} ]_{i,j=1,\dots,N} \frac{\det [ \lambda_i^{j-1}
]_{i,j=1,\dots,N} }{ \det [a_i^{j-1}]_{i,j=1,\dots,N} }.
\end{equation}
Simple manipulation and use of the Vandermonde determinant identity
reduces (\ref{S1}) in the case $t=1$ to  \eqref{eigenpdf} with
$V(x) = x^2$.

Now consider case 2). Let $\mathbf W$ be a $n \times N$ $(n \ge N)$
complex Gaussian matrix with real and imaginary parts having
variance 1/2, mean 0, and let $\mathbf B$ be an $N \times N$
positive definite Hermitian matrix with distinct eigenvalues
$b_1,\dots,b_N$. Again it is a straightforward consequence of
(\ref{IZ}) \cite{Baik2004,BK,GaoSmith,SimonMM} that the eigenvalue
p.d.f.~of
\begin{equation}\label{18.2}
\mathbf B^{-1/2} \mathbf W^\dagger \mathbf W \mathbf B^{-1/2}
\end{equation}
is equal to
\begin{equation}\label{S3}
\frac{(-1)^{N(N-1)/2} }{ N!} \prod_{j=1}^N \frac{b_j^N
\lambda_j^{n-N} }{ (n - N + j - 1)!} \prod_{1 \le i < j \le N}
\frac{(\lambda_i - \lambda_j) }{ (b_i - b_j) } \det [ e^{- b_i
\lambda_j} ]_{i,j=1,\dots,N}
\end{equation}
and thus with $b_i = 1+a_i$, $\lambda_j = x_j$, is proportional to
 \eqref{eigenpdf} with
$V(x) = x - (n-N) \log x$.

\begin{figure}[h]\caption{{\small
Contours $\mathcal{A}$ and
$\mathcal{A}_{\{s_1,\ldots,s_i\}}$.}}\label{FigAiry}
\begin{center}
 \begin{pspicture}(0,0)(6,6.5)
 \psline[linestyle=dashed]{-}(2,3)(4,6)\rput(4.3,6.3){{\small $\pi/3$}}
  \psline[linestyle=dashed]{-}(2,3)(4,0)\rput(4.3,-0.3){{\small $-\pi/3$}}
  \pscurve[linewidth=1.5pt]{->}(3.9,0)(2.4,1.8)(1.2,3)(2.4,4.2)(3.9,6)
  \rput(1,2){{\small $\mathcal{A}_{\{s_1,\ldots,s_i\}}$}}
  \pscurve[linewidth=1.5pt]{->}(4.1,0)(3,3)(4.1,6)
  \rput(3.5,4){{\small $\mathcal{A}$}}
  \psline{->}(0,3)(5.0,3)
     \rput(5.4,3){{\small $\Re (v)$}}
     \psline{->}(2,0)(2,6)
   \rput(2,6.3){{\small$\Im (v)$}}
      \rput(3.5,3.2){{$\cdots$}}
         \rput(1.5,3.2){{$s_1$}}\rput(2.5,3.2){{$s_2$}}\rput(4.5,3.2){{$s_i$}}
     \psdots*[dotstyle=*](1.5,3)(2.5,3)(4.5,3)
 \end{pspicture}
\end{center}
\end{figure}
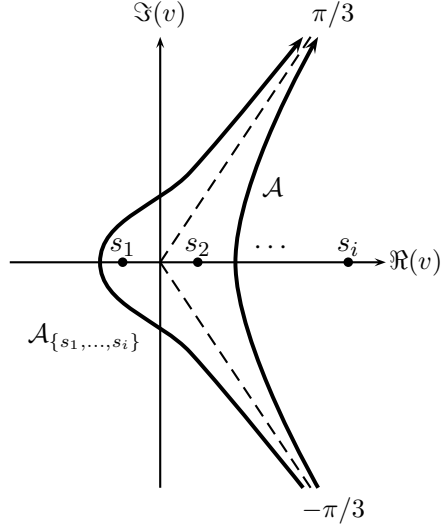

A specialization of (\ref{18.2}) of interest in mathematical
statistics \cite{Johnstone} is to take $\mathbf B = {\rm
diag}(b_1,\dots,b_r,\\ 1,\dots,1)$. Then the matrix $\mathbf X :=
 \mathbf W^\dagger \mathbf B^{-1/2}$ has columns $j$, $(j=1,\dots,r)$ with
variances $1/2\sqrt{b_j}$, while the entries in the remaining
columns all have variance $1/2$. Regarding $\mathbf X$ as a data
matrix, this corresponds to so called spiked data. It was shown in
\cite{Baik2004} that with the scaled variables $s_j$ and $X_j$
specified by
\begin{equation}\label{3.0}
b_j = \frac{1 }{ 2} + \frac{1}{2(2N)^{1/3}}{s_j} \: \: (j=1,\dots,r)
\qquad x_j =  4N + 2(2N)^{1/3} X_j \: \: (j=1,\dots,k)
\end{equation}
the scaled $k$-point correlation
\begin{equation}
\Big ( \frac{1}{2 (2N)^{1/3}}  \Big )^k \rho_{(k)}(x_1,\dots,x_k)
\end{equation}
tends in the limit $N \to \infty$, $n/N \to 1$ to
\begin{equation}\label{3.a}
\rho_{(k)}^{{\rm soft} \, \{s_j\}}(X_1,\dots,X_k) := \det \Big [
\mathcal{K}^{\mathrm{soft}}(X_\mu,X_\nu) \Big ]_{\mu,\nu=1, \dots,k}
\end{equation}
where
\begin{equation}\label{3.a1}
\mathcal{K}^{\mathrm{soft}}(X,Y)=K^{\mathrm{soft}}(X,Y)+\sum_{i=1}^r\widetilde{\mathrm{Ai}}^{(i)}(X)\mathrm{Ai}^{(i)}(Y)
\end{equation}
with
\begin{align}
K^{\mathrm{soft}}(X,Y)&=
\frac{\mathrm{Ai}(X)\mathrm{Ai}'(Y)-\mathrm{Ai}'(X)\mathrm{Ai}(Y)}{X-Y},
\label{3.a2} \\
\widetilde{\mathrm{Ai}}^{(i)}(x)&=\int_{\mathcal{A}_{\{s_1,\ldots,s_i\}}}\frac{dv}{2\pi\mathrm{i}}{e^{-xv+v^3/3}}{\prod_{k=1}^i(v-s_k)^{-1}},
\label{3.a3} \\
{\mathrm{Ai}}^{(i)}(x)&=(-1)^i\int_{\mathcal{A}}\frac{dv}{2\pi\mathrm{i}}{e^{-xv+v^3/3}}{\prod_{k=1}^{i-1}(v+s_k)}.
\label{3.a4}
 \end{align}
 See Fig.~\ref{FigAiry} for the definition of the contours $\mathcal{A}_{\{s_1,\ldots,s_i\}}$ and $\mathcal{A}$.
 (The same limit (\ref{3.a})
was found for $N \to \infty$ with $n/N \to \gamma^2$, $\gamma^2 \in
\mathbb R^+$, provided the scales in (\ref{3.0}) are suitably
modified.) The significance of the scale (\ref{3.0}) is that it
centres the coordinates about the neighbourhood of the largest
eigenvalue, and scales these eigenvalues so that their spacing is of
order unity. Because the eigenvalue density is not strictly zero
beyond the expected position of the largest eigenvalue, this is
referred to as the soft edge, and thus the reason for the
superscript `soft' in (\ref{3.a}) and (\ref{3.a1}).

Subsequent to the derivation of (\ref{3.a}) as a limiting
correlation at the soft edge for matrices (\ref{18.2}) with $\mathbf
B$ a rank $r$ perturbation of the identity, it was shown in
\cite{Peche} that (\ref{3.a}) is also the limiting correlation at
the soft edge for matrices (\ref{18.1}) with $\mathbf A$ a rank $r$
perturbation of the zero matrix. Explicitly, if in (\ref{18.1}) we
take $t=1$ and $\mathbf A = {\rm diag}(a_1,\dots,a_r,0,\dots,0)$,
with $a_1=\cdots = a_r =1$, and introduce the scaled variables $x_j
= \sqrt{2} N^{1/6}(\lambda_j - \sqrt{2N})$ $(j=1,\dots,k)$, then in
the limit $N \to \infty$ the scaled $k$-point correlation
$$
(\sqrt{2} N^{1/6})^k \rho_{(k)}(x_1,\dots,x_k)
$$
tends to (\ref{3.a}) with $s_1=\cdots=s_r=0$.

In both the studies \cite{Baik2004}, \cite{Peche} the structure
(\ref{3.a1}) appears after the asymptotic analysis of the respective
double contour representation of the general finite $N$, $(\mu,\nu)$
element in (\ref{3.a1}) (which is called the correlation kernel). It
is an objective of this to identity the structure analogous to
(\ref{3.a1}) in the finite $N$ p.d.f.~s (\ref{S1}), (\ref{S3}).
Moreover for (\ref{S3}) we identify a hard edge scaling (a scaling
of the smallest eigenvalues in (\ref{S3})) to the limiting $k$-point
correlation
\begin{equation}
\rho_{(k)}^{{\rm hard} \, \{h_j\}}(X_1,\dots,X_k) := \det \Big [
\mathcal{K}^{\mathrm{hard}}(X_\mu,X_\nu) \Big ]_{\mu,\nu=1, \dots,k}
\end{equation}
where
\begin{equation}\label{1.18}
\mathcal{K}^{\mathrm{hard}}(X,Y)=\left(\frac{Y}{X}
\right)^{(\alpha+r)/2}K^{\mathrm{hard}}_{\alpha+r}(X,Y)+
\sum_{i=1}^r\widetilde{J}^{(i)}(X){J}^{(i)}(Y)
\end{equation}
with
\begin{align}
K^{\mathrm{hard}}_\alpha(X,Y)&=\frac{J_\alpha(\sqrt{X})
\sqrt{Y}{J_\alpha}'(\sqrt{Y})-\sqrt{X}{J_\alpha}'(\sqrt{X})
J_\alpha(\sqrt{Y})}{2(X-Y)}, \label{3.b2}\\
\widetilde{J}^{(i)}(x)&=\int_{\mathcal{C}_{\{0,h_1,\ldots,h_i\}}}
\frac{dz}{2\pi\mathrm{i}}\frac{e^{-xz+1/4z}z^{\alpha+r}}
{\prod_{k=1}^i(z-h_k)},\label{3.b3} \\
{J}^{(i)}(x)&= \int_{\mathcal{C}_{\{0\}}}\frac{dw}{2\pi\mathrm{i}}
\frac{e^{xw-1/4w}\prod_{k=1}^{i-1}(w-h_k)}{w^{\alpha+r}}.\label{3.b4}
\end{align}
Here $\mathcal{C}_{\{0,h_1,\ldots,h_i\}}$ is a simple closed
anticlockwise contour encircling the points $h_i$ and the origin,
while $\mathcal{C}_{\{0\}}$ is a simple closed anticlockwise contour
encircling the origin (see Fig.\ \ref{FigC}).

The correlation functions obtained here are multiparameter
generalizations of the correlations  in the Hermite and Laguerre
unitary ensembles.  The latter quantities are reviewed in the next
section.  In \S3, we present some elements of the theory of multiple
orthogonal functions, which provides the natural setting for
studying the matrix ensembles with p.d.f.~of the form
\eqref{eigenpdf}.  We show in \S 4 that  $k$-point correlation
function can be written as $\det[K(x_i,x_j)]_{i,j=1,\ldots,k}$ for a
given kernel $K(x_i,x_j)$. The finite rank, $r$ say, perturbations
of the Hermite and Laguerre ensembles are considered in \S 5.  We
show that even for matrix ensembles of finite size $N$ the kernel
takes the form exhibited by \eqref{3.a1},  (\ref{1.18}). Explicitly
$K_N(x,y|\mathbf{a})=K_{N-r}(x,y|\mathbf{0})+\sum_{i=1}^r\tilde{f}_i(x)
f_i(y)$, where $\tilde{f}_i$ and  $f_i$ denote multiple functions
that generalize the Hermite or the Laguerre polynomials. In \S6, we
consider the large $N$ limit and obtain the scaled kernels
\eqref{3.a1}, \eqref{1.18}. This is achieved by showing that the
multiple Laguerre functions tend to multiple Bessel functions at the
hard edge, while both the multiple Laguerre and Hermite functions
tend to the same multiple Airy functions at the soft edge.  The
article ends with a short discussion.

\section{Null case}

The null case corresponds to
ensembles of random matrices with unitary symmetry (see for instance
\cite{ForresterBook, Metha}).   It is defined by
\begin{equation}
\mathbf{a}\longrightarrow\mathbf{0}=\{0,\ldots,0\}\end{equation} in
(\ref{eigenpdf}). In such a situation,
$\prod_{i<j}(a_j-a_i)\det\left[e^{-a_k x_l}\right]_{k,l}$ tends to
$\prod_{i<j}(x_j-x_i)$ , and the $k$-point correlation function can
be written as  a $k \times k$ determinant.

In order to be more concrete, let us introduce
$p_n(x)=x^n+c_{n-1}x^{n-1}+\ldots$, an orthogonal polynomial with
respect to the weight $w(x):=e^{-V(x)}$ supported on the real interval $J$;
that is,
\begin{equation}
\int_J dx\,w(x)p_j(x)p_k(x)=\|p_j  \|^2 \delta_{j,k}.
\end{equation}
Let also (Christoffel-Darboux formula \cite{Szego})
  \begin{equation}
\begin{split}\label{CD}
K_N(x,y)&:=\sqrt{w(x)w(y)}\sum_{n=0}^{N-1}\frac{p_n(x)p_n(y)}{\|p_n\|^2}\\
&=\frac{\sqrt{w(x)w(y)}}{\|p_{N-1}\|^2}\frac{p_N(x)p_{N-1}(y)-p_{N-1}(x)p_N(y)}{x-y}
\end{split}\end{equation} be the reproducing kernel on the interval $J$.  It
satisfies
\begin{equation}
\int_Jdy\, K_N(x,y)K_N(y,z)=K_N(x,z),\quad \int_Jdx\,K_N(x,x)=N
\end{equation}
and
\begin{equation}
P_N(
\mathbf{x}):=\frac{1}{Z_N}\prod_{i=1}^Nw(x_i)\prod_{1\leq j<k\leq
N}(x_k-x_j)^2 = \det
\left[K_N(x_i,x_j)\right]_{i,j=1}^N\,.\end{equation}
These properties together with the Laplace expansion of the
determinant imply that the $k$-point correlation function can be
expressed as a $k \times k$ determinant of the kernel,
\begin{equation} \label{CorrNullCase}
\begin{split}
\rho_{(k)}(x_1,\ldots,x_k)&:=\frac{N!}{(N-k)!}\int_J
dx_{k+1}\cdots \int_J dx_{N} P_N(x_1,\ldots,x_N)\\
&=\det \left[K_N(x_i,x_j)\right]_{i,j=1}^k\end{split}
\end{equation}

For the Hermite and Laguerre ensembles, the monic polynomials of
degree $n$ can  respectively be expressed in terms of the Hermite
and the Laguerre polynomials,
\begin{equation}
\begin{array}{llll}
1)& p_n(x)=2^{-n}H_n(x) & \| p_n\|^2=\sqrt{\pi}2^{-n}n!&
\mbox{if}\quad J=(-\infty,\infty) ,\\
2) &p_n(x)=(-1)^n n!L^\alpha_n(x) & \| p_n\|^2=n!(n+\alpha)!&
\mbox{if}\quad J=(0,\infty) .\\
\end{array}
\end{equation}
In the Hermite case, we
thus have
\begin{equation}\label{HermiteK}
K_N(x,y)=\frac{e^{-(x^2+y^2)/2}}{2^N\sqrt{\pi}(N-1)!}
\frac{H_N(x)H_{N-1}(y)-H_{N-1}(x)H_{N}(y)}{x-y},\end{equation} while
we have, in the Laguerre case,
\begin{equation}\label{LaguerreK}K^\alpha_N(x,y)=-\frac{N!}{(N+\alpha-1)!}(xy)^{\alpha/2}e^{-(x+y)/2}\frac{L_N^\alpha(x)L^\alpha_{N-1}(y)-L^\alpha_{N-1}(x)L^\alpha_{N}(y)}{x-y}.\end{equation}

Of particular interest are the scaled kernels \cite{ForresterEdge},
relating to the edge of the spectrum,
which are defined by
\begin{equation}\label{ScaledKernel}
K^{\mathrm{scaled}}(X,Y):=\lim_{N\rightarrow\infty}
B(N)K_N\left(A(N)+B(N)X,A(N)+B(N)Y\right), \end{equation}where
`scaled' stands for `hard' or `soft'.   When the
functions $A$ and $B$ are suitably chosen, one gets kernels that are
independent of $N$ and that can be expressed in terms of simple
functions.  In the Hermite ensemble, there is only one distinct
edge scaling,
\begin{equation}\label{HermiteEdges}x=A(N)+B(N)X=
\sqrt{2N}+X/\sqrt{2N^{1/3}}, \qquad \mathrm{soft\,
edge}.\end{equation}
Two edge scaling are possible for
the Laguerre ensemble,
\begin{equation}\label{LaguerreEdges}x=A(N)+B(N)X=\begin{cases}
X/4N,&\mathrm{hard\,edge},\\
4N+2(2N)^{1/3}X,&\mathrm{soft\, edge}.\end{cases}\end{equation} By
substituting the corresponding scaling of Eqs.\ \eqref{HermiteEdges} and
\eqref{LaguerreEdges} into Eq.~\eqref{ScaledKernel} the scaled
kernels (\ref{3.a2}), (\ref{3.b2}) result.

\section{Multiple biorthogonal polynomials}

\subsection{Generalities}

We   introduce the multiple  Hermite and Laguerre polynomials
\cite{Aptekarev, Bleher, BK}. Note that we use slightly different
notations and definitions than those of the latter references in
order to simplify the comparison between our asymptotic formulas and
the well known formulas as revised above in the null case.

Let us suppose that within the set 
\begin{equation}\label{aset}
\mathbf{a}=\{a_1,\ldots,a_N\},
\end{equation}
only $D$ eigenvalues are distinct. Then we may write
\begin{equation}\label{bset}
\mathbf{a}=\mathbf{b}^\mathbf{m}:=\{b_1^{m_1},\ldots,b_D^{m_D}\},
\end{equation}
which means  there are $m_i$  eigenvalues in $\mathbf{a}$ that are
equal to $b_i$ ($i=1,\dots,D$).   Let also
$|\mathbf{m}|:=\sum_{i=1}^D m_i=N$; that is, $\mathbf{m}$ is a
composition of non-negative integers, of weight $N$ and of fixed
length $D$.

The
$i$th multiple orthogonal polynomials of type I associated to a
given set of parameters $\mathbf{b}$, denoted by
$Q_{\mathbf{m},i}(x)$ or $Q_{\mathbf{m},i}(x|\mathbf{b})$ (but usually, the dependence on $\mathbf{b}$ is kept
implicit), is a polynomial in $x$ of degree $m_i-1$. It is used to
build the function $Q_\mathbf{m}(x):=\sum_{i=1}^D e^{-b_i
x}Q_{\mathbf{m},i}(x)$ such that
\begin{equation}
\int_J dx\, w(x)x^j Q_{\mathbf{m}}(x)=
\begin{cases}
0, &\quad j=0,\ldots,|\mathbf{m}|-2,\\
1,& \quad j=|\mathbf{m}|-1. \end{cases}\end{equation} The multiple
orthogonal polynomial of type II and of degree $|\mathbf{m}|$ is
denoted by $P_{\mathbf{m}}(x)$ or  $P_{\mathbf{m}}(x|\mathbf{b})$. It satisfies
\begin{equation}\int_Jdx\,w(x)e^{-b_ix} x^j P_{\mathbf{m}}(x)=0,\qquad
j=0,\ldots, m_i-1, \end{equation} where $i=1,\ldots,D$.  Combining
the two previous equations, we see that
\begin{equation}
\int_J dx\,w(x)P_{\mathbf{m}}(x)Q_\mathbf{n}(x)=
\begin{cases}
0, & |\mathbf{m}|<|\mathbf{n}|-1,\\
1, & |\mathbf{m}|=|\mathbf{n}|-1,\\
0, & \,m_i\, > \,n_i\,-1 \quad \forall i.
\end{cases}
\end{equation}
In other words, for a given set of parameters $\mathbf{b}$ and a
given composition $\mathbf{m}$ of weight $N$, the polynomials of
type I and II allow us to build a family of biorthogonal functions,
\begin{equation}
\int_J dx\, w(x)p_i(x)q_j(x)=\delta_{i,j}\qquad\mbox{for all}\quad
i,j=0,\ldots,N-1\end{equation}by choosing
\begin{equation}\label{3.4a}
p_i(x)=P_{\boldsymbol{\mu}_i}(x), \qquad
q_i(x)=Q_{\boldsymbol{\mu}_{i+1}}(x),
\end{equation}
and if for instance, the
compositions $\boldsymbol{\mu}_i$ are given by
\begin{equation}\label{3.4ap}
\left(\begin{array}{l}
                   \boldsymbol{\mu}_{N} \\
                      \boldsymbol{\mu}_{N-1} \\
                        \boldsymbol{\mu}_{N-2}\\
                         \boldsymbol{\mu}_{N-3}\\
                    \;\;\vdots \\
                    \boldsymbol{\mu}_{0}
                  \end{array}
                \right)=\left(
                          \begin{array}{llllll}
                            m_1& m_2 & m_3 & \ldots & m_D \\
                            m_1-1 & m_2 & m_3 & \ldots & m_D\\
                             m_1-1 & m_2-1 & m_3 & \ldots & m_D\\
                             m_1-1 & m_2-1& m_3-1 & \ldots & m_D\\
                            \;\;\vdots & \;\;\vdots &\;\vdots & \;\;\vdots &
  \;\vdots\\
                            \;\;0 & \;\;0 &\;0&  \cdots &  \;0\\
                          \end{array}
                        \right),
  \end{equation}
where it is understood that $m_i\geq m_{i+1}$.

\subsection{Multiple Laguerre functions}

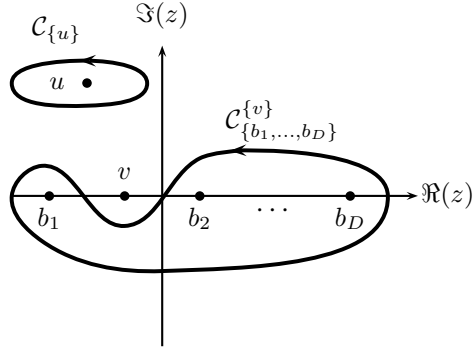
\begin{figure}[h]\caption{{\small
Contours $\mathcal{C}_{\{ u\}}$ and
$\mathcal{C}^{\{v\}}_{\{b_1,\ldots,b_D\}}$ in the complex
$z$-plane.}}\label{FigC}
\begin{center}
 \begin{pspicture}(0,0)(6,5)
\psdots*[dotstyle=*](1,3.5)\rput(0.6,3.5){{\small
$u$}}\rput(0.6,4.2){{\small $\mathcal{C}_{\{u\}}$}}
  \pscurve[linewidth=1.5pt]{->}(1,3.8)(0,3.5)(1,3.2)(1.8,3.5)(0.9,3.8)
  \pscurve[linewidth=1.5pt]{->}(3,2.6)(2.5,2.5)(1.5,1.6)(0.4,2.4)(0,2)(0.4,1.4)(2,1)(5,2)(2.9,2.6)
  \rput(3.6,3){{\small $\mathcal{C}^{\{v\}}_{\{b_1,\ldots,b_D\}}$}}
  \psline{->}(0,2)(5.4,2)
     \rput(5.8,2){{\small $\Re (z)$}}
     \psline{->}(2,0)(2,4)
   \rput(2,4.4){{\small$\Im (z)$}}
      \rput(3.5,1.8){{$\cdots$}}
         \rput(0.5,1.7){{\small $b_1$}}\rput(2.5,1.7){{\small $b_2$}}\rput(4.5,1.7){{\small $b_D$}}
         \rput(1.5,2.3){{\small $v$}}
     \psdots*[dotstyle=*](0.5,2)(1.5,2)(2.5,2)(4.5,2)
 \end{pspicture}
\end{center}
\end{figure}

The multiple Laguerre function of type I is \cite[
cf.~Eq.~(3.10)]{BK}
\begin{equation}\label{multiLaguerreI}
\widetilde{\mathcal{L}}^\alpha_\mathbf{m}(x)=\sum_{i=1}^De^{-b_i x}
\widetilde{\mathcal{L}}^\alpha_{\mathbf{m},i}(x):=
\int_{\mathcal{C}^{\{-1\}}_{\mathbf{b}}} \frac{dz}{2\pi\mathrm{i}}
\frac{e^{-xz}
(1+z)^{|\mathbf{m}|+\alpha-1}}{\prod_{i=1}^D(z-b_i)^{m_i}},\quad
\alpha\in\mathbb{Z},
\end{equation}
where
$\mathcal{C}^{\{-1\}}_{\mathbf{b}}=\mathcal{C}^{\{-1\}}_{\{b1,\ldots,b_D\}}$
denotes a closed contour which encircles positively $b_1,\ldots,b_D$
but not the point $-1$ (See Fig.~\ref{FigC}). Note that encircling
the point $-1$ is forbidden only when $\MM +\alpha-1$ is negative.
This means in particular that for $\alpha>-1$, the contour
$\mathcal{C}^{\{-1\}}_{\mathbf{b}}$ can be replaced by
$\mathcal{C}_{\mathbf{b}}$.  Obviously, the $i$th polynomial $
\widetilde{\mathcal{L}}^\alpha_{\mathbf{m},i}(x)$ has degree
$m_i-1$. The multiple Laguerre polynomial of type II has degree
$|m|$; it can be defined  by \cite[cf.~Eq.~(3.5)]{BK}
\begin{equation}\label{multiLaguerreII}
{\mathcal{L}}^\alpha_\mathbf{m}(x):=\frac{(|{\mathbf{m}}|+\alpha)!}{|\mathbf{m}|!}{x^{-\alpha}}\int_{\mathcal{C}_{\{0\}}}
\frac{dw}{2\pi \mathrm{i}}
\frac{e^{xw}}{w^{|\mathbf{m}|+\alpha+1}}\prod_{i=1}^D(w-1-b_i)^{m_i},\quad
\alpha\in\mathbb{Z},
\end{equation}
where $\mathcal{C}_{\{0\}}$ is a counterclockwise contour around the
origin. Our definitions, modifying those of \cite{BK}
 are motivated in part by the first of the
following results, while the key orthogonality type results remain
essentially unchanged.

\begin{proposition} \label{OrthoL} Let $L^\alpha_n(x)$ denote the Laguerre
polynomial of degree $n$.  Set $\alpha$ a non-negative integer and
$b_1,\ldots,b_D>-1$.  Then the multiple Laguerre functions satisfy
\[
\lim_{\mathbf{b}\rightarrow\mathbf{0}} \widetilde{\mathcal{L}}^
\alpha_\mathbf{m}(x)=L^\alpha_{|\mathbf{m}|-1}(x)\quad
\mbox{and}\quad
\lim_{\mathbf{b}\rightarrow\mathbf{0}}{\mathcal{L}}^\alpha_\mathbf{m}(x)=L^\alpha_{|\mathbf{m}|}(x).\]
Moreover \cite[Thm.~3.2]{BK}
\[\int_{0}^\infty
dx\,  e^{-x}x^{\alpha+j} \widetilde{\mathcal{L}}^\alpha_\mathbf{m}(x)=
\begin{cases}0,&j=0,\ldots,|\mathbf{m}|-2\\
\displaystyle
\frac{(-1)^{|\mathbf{m}|-1}(|\mathbf{m}|+\alpha-1)!}{\prod_{i=1}^D(1+b_i)^{m_i}},&j=|\mathbf{m}|-1,\end{cases}\]
and  \cite[Thm.~3.1]{BK}
\[ \int_{0}^\infty
dx\, e^{-x-b_ix}x^{\alpha+j} {\mathcal{L}}^\alpha_\mathbf{m}(x)=0
\quad\mbox{for}\quad i=1,\ldots,D\quad\mbox{and}\quad j=0,\ldots,m_i-1.\]
\end{proposition}
\begin{proof}One easily proves the two first properties by considering the
integral representation of the Laguerre polynomials \cite{Szego},
\begin{equation}\label{intrepHermite}
L^\alpha_n(x)=\int_{\mathcal{C}_{\{0\}}}\frac{dw}{2\pi\mathrm{i}}
\frac{e^{-xw}}{w^{n+1}}(1+w)^{n+\alpha} \,\end{equation} and the
readily verified formula
\begin{equation}\label{3.8a}
n!(-x)^{-\alpha} L^{-\alpha}_n(x)=(n-\alpha)!L^\alpha_{n-\alpha}(x),
\end{equation}
both relations
being valid for $\alpha\in\mathbb{Z}$. Then, using
Eq.~\eqref{multiLaguerreI}, we have
\[\begin{split}
\int_{0}^\infty dx\,  e^{-x}x^{\alpha+j}
\widetilde{\mathcal{L}}^\alpha_\mathbf{m}(x)&=\int_{\mathcal{C}_{\mathbf{b}}}
\frac{dz}{2\pi\mathrm{i}}
\frac{(1+z)^{|m|+\alpha-1}}{\prod_{i=1}^D(z-b_i)^{m_i}}\int_{0}^\infty
dx\,  e^{-x-xz}x^{\alpha+j}\\
&=(\alpha+j)!\int_{{\mathcal{C}^{\{-1\}}_{\mathbf{b}}}}
\frac{dz}{2\pi\mathrm{i}}
\frac{(1+z)^{|m|-j-2}}{\prod_{i=1}^D(z-b_i)^{m_i}}
\end{split}\]
Note that, on the first line,  we assumed $b_i >-1$ and $\Re (z)>-1$;
this allowed us to do the $x$-integration. For $j=|\mathbf{m}|-1$,
we use the residue theorem and get
\[\int_{0}^\infty
dx\,  e^{-x}x^{\alpha+j}
\widetilde{\mathcal{L}}^\alpha_\mathbf{m}(x)=-(|\mathbf{m}|+\alpha-1)!
\mathrm{Res}_{z={-1}}
\left[\prod_{i=1}^D(z-b_i)^{-m_i}\right]=-\frac{(|\mathbf{m}|+\alpha-1)!}{\prod_{i=1}^D(-1-b_i)^{m_i}}\]as
desired. For $j<|\mathbf{m}|-1$, we consider
$(1+z)^{|m|-j-2}\prod_{i=1}^D(z-b_i)^{-m_i}\rightarrow z^{-j-2}$
when $z\rightarrow\infty$.  Thus, by virtue of the residue theorem,
\[\int_{0}^\infty
dx\,  e^{-x}x^{\alpha+j}
\widetilde{\mathcal{L}}^\alpha_\mathbf{m}(x)\propto
\mathrm{Res}_{z={\infty}}\left[z^{-j-2}\right] =0 \quad\mbox{when}\quad
-1<j<|\mathbf{m}|-1
.\]The same tricks are used to prove the second orthogonality
property.  Explicitly,
\[\begin{split}
\int_{0}^\infty dx\, e^{-x-b_ix}x^{\alpha+j}
{\mathcal{L}}^\alpha_\mathbf{m}(x)&\propto
\int_{\substack{\mathcal{C}_{\{0\}}\\ \Re (w)<1+b_i}} \frac{dw}{2\pi
\mathrm{i}}
\frac{\prod_{j=1}^D(w-1-b_j)^{m_j}}{w^{|\mathbf{m}|+\alpha+1}}\int_{0}^\infty
dx\, e^{-x-b_ix+xw}x^{j}\\
&\propto \int_{\mathcal{C}_{\{0\}}} \frac{dw}{2\pi \mathrm{i}}
\frac{(w-1-b_i)^{-j-1}\prod_{j=1}^D(w-1-b_j)^{m_j}}{w^{|\mathbf{m}|+\alpha+1}}\\
&\propto
  \mathrm{Res}_{w={\infty}}\left[w^{-j-\alpha-2}\right]=0.\\
\end{split}\]when $-1<j<m_i$ and $\alpha>-1$, as proposed.
\end{proof}

Comparing with the general formalism of \S 3.1 we have
\begin{eqnarray}\label{wPQ}
w(x) & =  & x^\alpha e^{-x} \nonumber \\
Q_{\mathbf{m}}(x) & = & \frac{(-1)^{|\mathbf{m}|-1} \prod_{i=1}^D (1
+ b_i)^{m_i} }{ (\mathbf{m} + \alpha - 1)!}
\tilde{\mathcal L}_{\mathbf{m}}^\alpha(x) \nonumber \\
P_{\mathbf{m}}(x) & = & {\mathcal L}_{\mathbf{m}}^\alpha(x)
\end{eqnarray}
with corresponding biorthogonal functions specified by (\ref{3.4a}).

The multiple Laguerre functions also satisfy certain differential
equations \cite{Aptekarev}.  Indeed, if
\begin{equation}
\label{opD}\mathcal{D}_{t,x}f(x):=-e^{tx}\partial_x\left(e^{-tx}f(x)\right)=(t-\partial_x)f(x), \quad \mbox{with}\quad\partial_x=\frac{\partial}{\partial x},
\end{equation}
then
\begin{equation}\label{3.9'}
\mathcal{D}_{t,x}\left[e^{-xz}\right]=(z+t)e^{-xz}.
\end{equation}
 Consequently,
 \begin{equation}
0
= \mathcal{D}_{-a_1,x}\cdots\mathcal{D}_{-a_{\MM},x}\left[\LLt(x)\right]
\end{equation}
 and
 \begin{equation}
x^\alpha\LL(x)
=\frac{(-1)^{\MM}}{\MM!}
 \mathcal{D}_{1+a_1,x}\cdots\mathcal{D}_{1+a_{\MM},x}\left[x^{\MM+\alpha}\right].
\end{equation}
When $m_D$ parameters in $\aaa$ are equal to zero, $a_{\MM-m_D}=\ldots=a_{\MM}=0$ say,
after some simple
manipulations we can show that the multiple Laguerre functions of type I and II can be respectively
interpreted as anti-derivatives and derivatives
of Laguerre polynomials.  Explicitly, for $r=\MM-m_D$, we see from the
integral representations (\ref{multiLaguerreI}), (\ref{intrepHermite})
and the differentiation formula (\ref{3.9'}) that
\begin{equation}\label{3.12}
L^{\alpha+r}_{\MM-r-1}=
\mathcal{D}_{-a_1,x}\cdots\mathcal{D}_{-a_{r},x}\left[\LLt(x)\right]
\end{equation}
while the integral representations (\ref{multiLaguerreI}), (\ref{intrepHermite}), the identity (\ref{3.9'}) and the differentiation formula (\ref{3.9'}) show that
\begin{equation}\label{3.13}
x^\alpha\LL(x)=(-1)^r\frac{(\MM-r)!}{\MM!} \mathcal{D}_{a_1,x}\cdots\mathcal{D}_{a_r,x}\left[x^{\alpha+r}L^{\alpha+r}_{\MM-r}(x)\right].
\end{equation}

\subsection{Multiple Hermite functions}

The multiple Hermite polynomials of type I, written  $
\widetilde{\mathcal{H}}_{\mathbf{m},i}(x)$,   are obtained  from the
function \cite[cf.~({2.9})]{BK}
\begin{equation}\label{multiHermiteI}
\widetilde{\mathcal{H}}_\mathbf{m}(x)=\sum_{i=1}^De^{-b_i x}
\widetilde{\mathcal{H}}_{\mathbf{m},i}(x)=(-2)^{|\mathbf{m}|-1}(|\mathbf{m}|-1)!\int_{\mathcal{C}_\mathbf{b}}\frac{dv}{2\pi\mathrm{i}}\frac{e^{-v^2/4-xv}}{\prod_{i=1}^D(v-b_i)^{m_i}},\end{equation}
where $\mathcal{C}_\mathbf{b}$ is a  simple contour which encircles
the points $\{b_1,\ldots,b_D\}$ anticlockwise. The (non-monic)
multiple Hermite polynomial of type II is given by
\cite[cf.~({2.2})]{BK}
\begin{equation}\label{multiHermiteII}
{\mathcal{H}}_\mathbf{m}(x)=(-1)^{|\mathbf{m}|}\sqrt{\pi}e^{x^2}\int_{-\mathrm{i}\infty}^{\mathrm{i}\infty}\frac{du}{2\pi\mathrm{i}}e^{u^2/4+xu}\prod_{i=1}^D(u-b_i)^{m_i}\,.\end{equation}
One can verify that the degrees of $
\widetilde{\mathcal{H}}_{\mathbf{m},i}(x)$ and
${\mathcal{H}}_\mathbf{m}(x)$ are respectively $m_i-1$ and
$|\mathbf{m}|$.

The multiple Hermite functions (polynomials) can be obtained as a limit
of their Laguerre counterpart.

\begin{proposition}\label{LtoH} Let $X$  be real numbers such that
$|\sqrt{2\alpha}X|<\alpha$ with $\alpha\in\mathbb{N}$.  Then, as
$\alpha\rightarrow\infty$,
\[(2\alpha)^{(1-\MM)/2}\LLt(\alpha+\sqrt{2\alpha}X)\Big|_{b_i\mapsto
b_i/\sqrt{2\alpha}}=\frac{(-1)^{\MM-1}}{2^{\MM-1}(\MM-1)!}\HHt(X)+\order{\frac{1}{\sqrt{\alpha}}}\]and
\[(2\alpha)^{-\MM/2}\LL(\alpha+\sqrt{2\alpha}X)\Big|_{b_i\mapsto
b_i/\sqrt{2\alpha}}=\frac{(-1)^{\MM}}{2^{\MM}\MM!}\HH(X)+\order{\frac{1}{\sqrt{\alpha}}}.\]
\end{proposition}
\begin{proof}
The  proof of the first asymptotic formula is the simplest.  From
the definition of the type I Laguerre functions we have
\[(2\alpha)^{(1-\MM)/2}\LLt(\alpha+\sqrt{2\alpha}X)\Big|_{b_i\mapsto
b_i/\sqrt{2\alpha}}=(2\alpha)^{(1-\MM)/2}\int_{\mathcal{C}_{\mathbf{b}/\sqrt{2\alpha}}}
\frac{dz}{2\pi\mathrm{i}} \frac{e^{-\alpha z-\sqrt{2\alpha}Xz}
(1+z)^{|m|+\alpha-1}}{\prod_{i=1}^D(z-b_i/\sqrt{2\alpha})^{m_i}}.\]By
changing $z \mapsto  z/\sqrt{2\alpha}$, we get
\[(2\alpha)^{(1-\MM)/2}\LLt(\alpha+\sqrt{2\alpha}X)\Big|_{b_i\mapsto
b_i/\sqrt{2\alpha}}=\int_{\mathcal{C}_{\mathbf{b}}}
\frac{dz}{2\pi\mathrm{i}} \frac{e^{-X z-\sqrt{\alpha/2}z}
(1+z/\sqrt{2\alpha})^{\alpha+|m|-1}}{\prod_{i=1}^D(z-b_i)^{m_i}}.\]From
the binomial development $(1+z/\sqrt{2\alpha})^{\alpha+|m|-1}$ and
the Maclaurin expansion of $e^{-\sqrt{\alpha/2}z}$,  we easily find
that
\[\begin{split}e^{-\sqrt{\alpha/2}z}(1+z/\sqrt{2\alpha})^{\alpha+|m|-1}
&=e^{-\sqrt{\alpha/2}z}(1+z/\sqrt{2\alpha})^{\alpha}+\order{\frac{1}{\sqrt{\alpha}}}\\
&=e^{-z^2/4}+\order{\frac{1}{\sqrt{\alpha}}},\end{split}\] when
$\alpha\rightarrow\infty$.  This immediately implies the first
asymptotic formula.

The second asymptotic relation is established by steepest descents.
The integral formula (\ref{multiLaguerreII}) shows
\[
\LL(\alpha+\sqrt{2\alpha}X)=\frac{(|{\mathbf{m}}|+\alpha)!}{|\mathbf{m}|!}(\alpha+\sqrt{2\alpha}X)^{-\alpha}\int_{\mathcal{C}_{\{0\}}}
\frac{dw}{2\pi \mathrm{i}} e^{\alpha f(w)}g(w),\] where
\[f(w)=w-\ln w \quad\mbox{and}\quad
g(w)={e^{-\alpha\ln(1+\sqrt{2/\alpha}X)}}e^{\sqrt{2\alpha}X w}
\prod_{i=1}^D(w-1-b_i)^{m_i}{w^{-|\mathbf{m}|-1}}.
\]
We point out that $\alpha w$ dominates $\sqrt{2\alpha}X w$  by
hypothesis.  Thus, the function $f(w)$ determines the principal
contribution to the integral when $\alpha$ becomes large.   $f(w)$
possesses one simple saddle point at $w_0=1$ ; that is, $f'(w_0)=0$
but $f''(w_0)=1\neq 0$.  The steepest descent contour is fixed by
the condition $(w-w_0)^2 f''(w_0)=\mbox{minimum}$, which means
$\mathrm{arg}(w)=\pm \pi/2$.  We consequently choose the contour
$\mathcal{C}_0$ of $w$ such that the latter variable starts at
$1-\mathrm{i}c$ ($c$ is real and positive), passes through the point
$w=1$, goes to $1+\mathrm{i}c$ and finally returns to the starting
point after encircling that origin.
We now set $v=\sqrt{2\alpha}(w-1)$. When
$\alpha\rightarrow\infty$
\[\begin{split}
&\alpha
f(w)=\alpha+v^2/4+\order{\frac{1}{\sqrt{\alpha}}};\\
& (2\alpha)^{\MM/2}g(w)= e^{X
v+X^2}\prod_{i=1}^D(v-\sqrt{2\alpha}b_i)+\order{\frac{1}{\sqrt{\alpha}}};
\end{split}\]
and the steepest descent path in the complex $v$-plane becomes the
interval $(-\mathrm{i}\infty, \mathrm{i}\infty)$ on the imaginary
axis.  We also know from Stirling's approximation that
\[(|{\mathbf{m}}|+\alpha)!=\sqrt{2\pi}e^{-\alpha}\alpha^{\MM
+\alpha+1/2}+\order{\frac{1}{{\alpha}}}.\]Therefore, we have proved
\[(2\alpha)^{-\MM/2}\LL(\alpha+\sqrt{2\alpha}X)\Big|_{b_i\mapsto
b_i/\sqrt{2\alpha}}=\frac{1}{2^{\MM}\MM!}
\sqrt{\pi}e^{X^2}\int_{-\mathrm{i}\infty}^{\mathrm{i}\infty}\frac{dv}{2\pi\mathrm{i}}e^{v^2/4+xu}\prod_{i=1}^D(v-b_i)^{m_i},\]
which is the desired expression.
\end{proof}

\begin{proposition}  Let $H_n(x)$ be the Hermite polynomial of degree $n$.
Then the multiple
Hermite functions satisfy
\[
\lim_{\mathbf{b}\rightarrow\mathbf{0}}
\widetilde{\mathcal{H}}_\mathbf{m}(x)=H_{|\mathbf{m}|-1}(x)\quad
\mbox{and}\quad
\lim_{\mathbf{b}\rightarrow\mathbf{0}}{\mathcal{H}}_\mathbf{m}(x)=H_{|\mathbf{m}|}(x).\]
Moreover \cite[Thm 2.3]{BK}\[\int_{-\infty}^\infty dx\, e^{-x^2}x^j
\widetilde{\mathcal{H}}_\mathbf{m}(x)=
\begin{cases}0,&j=0,\ldots,|\mathbf{m}|-2\\
\displaystyle
\sqrt{\pi}(|\mathbf{m}|-1)!,&j=|\mathbf{m}|-1,\end{cases}\] and
\cite[Thm 2.1]{BK}\[ \int_{-\infty}^\infty dx\, e^{-x^2-b_ix}x^j
{\mathcal{H}}_\mathbf{m}(x)=0 \quad\mbox{for}\quad
i=1,\ldots,D\quad\mbox{and}\quad j=0,\ldots,m_i-1.\]
\end{proposition}
\begin{proof}The two first properties are straightforward
consequences of well known integral representation of the Hermite
polynomials \cite{Szego}
\begin{equation}\label{intrepHermite}
H_n(x)=2^nn!\int_{\mathcal{C}_{\{0\}}}\frac{dw}{2\pi\mathrm{i}}\frac{e^{xw-w^2/4}}{w^{n+1}}=
\sqrt{\pi}e^{x^2}\int_{-\mathrm{i}\infty}^{\mathrm{i}\infty}\frac{dw}{2\pi\mathrm{i}}w^ne^{-xw+w^2/4}.\end{equation}
We can prove the orthogonality relations by using Proposition
\ref{LtoH} and substituting
\[
e^{-\sqrt{2\alpha}x}(1+\sqrt{\frac{2}{\alpha}}x)^{\alpha}=e^{-x^2}+\order{\frac{1}{\sqrt{\alpha}}}\]
in the last equations of Proposition \ref{OrthoL}.
\end{proof}

Suppose in (\ref{multiHermiteI}) and (\ref{multiHermiteII}) that
$b_i = a_i$, $m_i = 1$ $(i=1,\dots,r)$ while $b_{r+1}=0$, $m_{r+1} =
|\mathbf{m}|-r$. Then we see from (\ref{3.9'}) and
(\ref{intrepHermite}) that
\begin{equation}\label{m.1}
{\mathcal D}_{-a_1,x} \cdots {\mathcal D}_{-a_r,x}
\widetilde{\mathcal H}_{\mathbf{m}}(x) = (-2)^r \frac{(|\mathbf{m}|
- 1)! }{ (|\mathbf{m}| - r - 1)! } H_{|\mathbf{m}| - r - 1}(x)
\end{equation}
(cf.~(\ref{3.12})) and
\begin{equation}\label{m.2}
{\mathcal D}_{a_1,x} \cdots {\mathcal D}_{a_r,x} \Big ( e^{-x^2}
H_{|\mathbf{m}| - r}(x) \Big ) = e^{-x^2}  \widetilde{\mathcal
H}_{\mathbf{m}}(x)
\end{equation}
\section{Non-null case}

\subsection{Multiple Laguerre kernel}

For the determinantal point processes (\ref{DP}), the method of
biorthogonal polynomials \cite{Borodin} gives that the corresponding
$k$-point correlation function has the determinant form
\begin{equation}\label{CorrDet}
\rho_{(k)}(x_1,\dots,x_k) = \det [{\mathcal K}(x_i,x_j) ]_{i,j=1,\dots,k}
\end{equation}
for a certain ${\mathcal K}(x,y)$ referred to as the correlation
kernel. The correlation kernel can always be written as a double sum
over $\xi_i(x) \eta_j(y)$ weighted by a factor proportional to the
inverse of the matrix of inner products $[(\xi_i,\eta_j)]$. We know
from \cite{Baik2004} that for the Laguerre case of (\ref{DP}) this
double sum can be written as a double contour integral
\begin{equation}\label{multiLK}
\KKL(x,y)=\left(\frac{x}{y}\right)^{\alpha/2}e^{(y-x)/2}
\int_{\mathcal{C}^{\{-1\}}_\mathbf{b}}\frac{dz}{2\pi\mathrm{i}}\int_{\mathcal{C}_{\{-1\}}}\frac{dw}{2\pi\mathrm{i}}
\frac{e^{-xz+yw}}{w-z}\left(\frac{1+z}{1+w}\right)^{\MM+\alpha}
\prod_{i=1}^D\left(\frac{w-b_i}{z-b_i}\right)^{m_i},
\end{equation}
In the latter expression, $\alpha$ stands for a non-negative integer
while $\mathcal{C}^{\{-1\}}_\mathbf{b}$ and $\mathcal{C}_{\{-1\}}$
stand for {\it non-intersecting} counterclockwise contours that
encircle $\bb$ and $-1$, respectively.

On the other hand, a determinantal point process
\begin{equation}\label{4.p1}
\frac{1 }{ Z_N} \prod_{l=1}^N w(x_l) \det [ p_i(x_j)
]_{i,j=1,\dots,N} \det [ q_i(x_j) ]_{i,j=1,\dots,N}
\end{equation}
in which $\{p_i(x)\}$, $\{q_j(x)\}$ have the biorthogonality property,
has for its correlation kernel the single sum
\begin{equation}\label{4.p2}
(w(x) w(y))^{1/2} \sum_{i=1}^N p_i(x) q_i(y)
\end{equation}
(cf.~the first line in (\ref{CD})). Furthermore, the general theory
of multiple orthogonal polynomials \cite{DK} tells us that with
$p_i, q_j$ as in (\ref{3.4a}), (\ref{3.4ap}) can be summed according
to a generalization of the Christoffel-Darboux formula in
(\ref{CD}). In the Laguerre case this summation has been made
explicit in \cite{BK}. We revise this latter formula, and make
explicit the form of (\ref{4.p2}) in the following.

\begin{proposition}\label{DarbouxL}
Let $\mathbf{e}_i=(\overbrace{0,\ldots,0}^{i-1},1,0\ldots)$.  The
correlation kernel (\ref{multiLK}) has the summed series form
\[
\KKL(x,y)=\frac{\MM!}{(\MM+\alpha-1)!}\frac{(xy)^{\alpha/2}
e^{-(x+y)/2}}{x-y}\left( \LLt(x)\LL(y)-\sum_{i=1}^D\frac{m_i}{\MM}
\widetilde{\mathcal{L}}^\alpha_{\mathbf{m}+\mathbf{e}_i}(x){\mathcal{L}}^\alpha_{\mathbf{m}-\mathbf{e}_i}(y)\right)\]
and the single sum form
\[\KKL(x,y)={(xy)^{\alpha/2}
e^{-(x+y)/2}}\sum_{i=1}^D\sum_{j=1}^{m_i}\frac{(\MM-j)!}{(\MM+\alpha-j)!}
e^{-b_ix}\widetilde{\mathcal{L}}^\alpha_{\mathbf{m}-(j-1)\mathbf{e}_i,i}(x){\mathcal{L}}^\alpha_{\mathbf{m}-j\mathbf{e}_i}(y).\]
\end{proposition}
\begin{proof} First,  we effectively
multiply Eq.~\eqref{multiLK} by $(x-y)/(x-y)$ by writing
\begin{multline*}\KKL(x,y)=\left(\frac{x}{y}\right)^{\alpha/2}\frac{e^{(y-x)/2}}{x-y}\\
\times
\int_{\mathcal{C}^{\{-1\}}_\mathbf{b}}\frac{dz}{2\pi\mathrm{i}}\int_{\mathcal{C}_{\{-1\}}}\frac{dw}{2\pi\mathrm{i}}
\frac{1}{w-z}\left(\frac{1+z}{1+w}\right)^{N+\alpha}
\prod_{i=1}^N\left(\frac{w-a_i}{z-a_i}\right)\left(-\frac{\partial}{\partial
z}-\frac{\partial}{\partial w}\right)(e^{-xz+yw}),\end{multline*}
where use has been made of the relation between $\{b_i\}$ and
$\{a_i\}$ noted in \S 3.1. Integrating by parts gives
\begin{multline*}\KKL(x,y)=\left(\frac{x}{y}\right)^{\alpha/2}\frac{e^{(y-x)/2}}{x-y}\\
\times
\int_{\mathcal{C}^{\{-1\}}_\mathbf{b}}\frac{dz}{2\pi\mathrm{i}}\int_{\mathcal{C}_{\{-1\}}}\frac{dw}{2\pi\mathrm{i}}
{e^{-xz+yw}}\left(\frac{1+z}{1+w}\right)^{N+\alpha}
\prod_{i=1}^N\left(\frac{w-a_i}{z-a_i}\right)\left(\frac{N+a}{(1+z)(1+w)}-\sum_{i=1}^N\frac{1}{(z-a_i)(w-a_i)}\right)\end{multline*}
which which from (\ref{multiLaguerreI}) is equivalent to the first
equation of the proposition.

Second, we apply Cauchy's integral theorem to Eq.~\eqref{multiLK}
and obtain
\begin{multline*}\KKL(x,y)=\left(\frac{x}{y}\right)^{\alpha/2}{e^{(y-x)/2}}\sum_{i=1}^D\int_{\mathcal{C}^{\{-1\}}_{\{b_i\}}}\frac{dz}{2\pi\mathrm{i}}\frac{e^{-xz}(1+z)^{\MM+\alpha}}{\prod_{i=1}^D\left({z-b_i}\right)^{m_i}}\\
\times \int_{\mathcal{C}_{\{-1\}}}\frac{dw}{2\pi\mathrm{i}}
\frac{e^{yw}}{(1+w)^{\MM+\alpha}}
\prod_{i=1}^D\left({w-b_i}\right)^{m_i}\frac{1}{w-z}\end{multline*}
We can choose  the contours $\mathcal{C}^{\{-1\}}_\mathbf{b_i}$ and
$\mathcal{C}_{\{-1\}}$ in a manner that guarantees
$|z-b_i|/|w-b_i|<1$ and $|1+w|/|1+z|<1$. In such a situation, we
have
\[\frac{1}{w-z}=\sum_{n=0}^\infty
\frac{(z-b_i)^n(1+w)^n}{(w-b_i)^{n+1}(1+z)^{n+1}} ,\] and
consequently,
\begin{multline*}\KKL(x,y)=\left(\frac{x}{y}\right)^{\alpha/2}{e^{(y-x)/2}}\sum_{i=1}^D\sum_{j=0}^{\infty}\int_{\mathcal{C}^{\{-1 \}}_{\{b_i\}}}\frac{dz}{2\pi\mathrm{i}}\frac{e^{-xz}(1+z)^{\MM+\alpha-j-1}}{(z-b_i)^{m_i-j}\prod_{k\neq
i}\left({z-b_k}\right)^{m_k}}\\
\times \int_{\mathcal{C}_{\{-1\}}}\frac{dw}{2\pi\mathrm{i}}
\frac{e^{yw}}{(1+w)^{\MM+\alpha-j}}(w-b_i)^{m_i-j-1} \prod_{j\neq
i}\left({w-b_k}\right)^{m_k}\end{multline*} The integral in $z$
around $b_i$ has no pole when $j\geq m_i$ and so the sum over $j$
can be truncated.  We thus can write
\begin{multline*}\KKL(x,y)=\left({x}{y}\right)^{\alpha/2}{e^{-(x+y)/2}}\sum_{i=1}^D\sum_{j=0}^{m_j-1}\int_{\mathcal{C}_{\{b_i\}}}\frac{dz}{2\pi\mathrm{i}}\frac{e^{-xz}(1+z)^{\MM+\alpha-j-1}}{(z-b_i)^{m_i-j}\prod_{k\neq
i}\left({z-b_k}\right)^{m_k}}\\
\times
y^{-\alpha}e^{-y}\int_{\mathcal{C}_{\{-1\}}}\frac{dw}{2\pi\mathrm{i}}
\frac{e^{yw}}{(1+w)^{\MM+\alpha-j}}(w-b_i)^{m_i-j-1} \prod_{j\neq
i}\left({w-b_k}\right)^{m_k}.\end{multline*} Simple manipulations
and recalling (\ref{multiLaguerreI}) complete the proof.\end{proof}

We now turn our attention to the correlation functions. It is known from
\cite{Baik2004} that they are given by \eqref{CorrDet} with
correlation kernel \eqref{multiLK}. To give further insight into
this result, we will re-establish this fact using a formalism
analogous to that revised in \S 2 for the null case. In particular
we want to prove Eq.~\eqref{CorrNullCase} remains valid in the
non-null case. According to the discussion of \S1 the perturbed
Laguerre p.d.f.~is given by
\begin{equation}\label{pdfL}
P_N^\alpha(x_1,\dots,x_N)=\frac{1}{Z^\alpha_N}\prod_{i=1}^Nx_i^\alpha
e^{-x_i}\prod_{1\leq i<j\leq
N}\frac{x_j-x_i}{a_j-a_i}\det\left[e^{-a_ix_j}\right]_{i,j=1}^N,
\end{equation} where it is supposed that no eigenvalues in
$\aaa$ coincide.
When $a_i=a_j$ for some $i\neq j$, the appropriate  p.d.f.~is obtained by
applying L'Hospital's rule on $P_N^\alpha(x)$.
Further, we read off from (\ref{S3}) that
\begin{equation}\label{lemmaZalphaN}
Z^\alpha_N=(-1)^{N(N-1)/2}N!\prod_{i=1}^N(\alpha+i-1)!
\prod_{i=1}^N(1+a_i)^{-N-\alpha}.
\end{equation}

\begin{lemma}\label{lemmadetKLP}
We have
$\displaystyle
\det\left[\KKL(x_i,x_j)\right]_{i,j=1}^{\MM}=\MM!P^\alpha_{\MM}(x_1,\ldots,x_{\MM})$.\end{lemma}
\begin{proof}
Let us
suppose $a_1<\ldots<a_N$ where $N=\MM$; i.e., $\mm=(1,\ldots,1)$.
Then Cauchy's theorem allows us to expand Eq.~\eqref{multiLK},
\[\begin{split}
\KKL(x,y)
&=\left(\frac{x}{y}\right)^{\alpha/2}e^{(y-x)/2}\sum_{i=1}^N
\int_{\mathcal{C}_{\{-1\}}}\frac{dw}{2\pi\mathrm{i}}
e^{-xa_i+yw}\left(\frac{1+a_i}{1+w}\right)^{N+\alpha}
\prod_{\substack{j=1\\j\neq i}}^N\left(\frac{w-a_j}{a_i-a_j}\right)\\
&=\left(\frac{x}{y}\right)^{\alpha/2}e^{-(x+y)/2}\sum_{i=1}^N
\frac{\left({1+a_i}\right)^{\MM+\alpha}e^{-xa_i}}
{\prod_{\substack{k=1\\k\neq
i}}^N\left({a_i-a_k}\right)}\int_{\mathcal{C}_{\{0\}}}\frac{dw}{2\pi\mathrm{i}}\frac{e^{yw}}{w^{N+\alpha}}
\prod_{\substack{j=1\\j\neq i}}^N\left({w-1-a_j}\right).
\end{split}\]
But we have
\[\begin{split}
\int_{\mathcal{C}_{\{0\}}}\frac{dw}{2\pi\mathrm{i}}\frac{e^{yw}}{w^{n}}&=\frac{y^{n-1}}{(n-1)!},\\
\prod_{\substack{j=1\\j\neq
i}}^N\left({w-1-a_j}\right)&=\sum_{j=1}^N(-1)^{N-j}e^{(i)}_{N-j}(\aaa')w^{j-1},
\end{split}\]
where $\aaa'=\mathbf{1+a}=(1+a_1,\ldots,1+a_N)$ and where
$e^{(i)}_n(\mathbf{x})$ stands for the elementary symmetric function
of degree $n$ that does not contain the variable $x_i$ \cite[\S
I.3]{Macdonald}. Combining the last equations we get
\begin{equation}
\label{KLcij}\KKL(x,y)=(xy)^{\alpha/2}e^{-(x+y)/2}\sum_{i,j=1}^Ne^{-a_ix}C_{i,j}y^{j-1}
\end{equation}
where
\begin{equation}
\label{cij}
C_{i,j}=\frac{(-1)^{j-1}(1+a_i)^{N+\alpha}}{(\alpha+j-1)!}e^{(i)}_{j-1}(\aaa'){\prod_{\substack{k=1\\k\neq
i}}^N\left({a_i-a_k}\right)^{-1}}.
\end{equation}
We finally use
\[\prod_{i<j}(a_j-a_i)=\prod_{i<j}(a_j'-a_i')=\det\left[(a_i')^{j-1}\right]_{i,j}
=\det\left[e^{(i)}_{j-1}(\aaa')\right]_{i,j}\] and
the general formula
\[
\det\left[\sum_{k,l}c_{k,l}f_k(x_i)g_l(x_j)\right]_{i,j}=
\det\left[f_i(x_j)\right]_{i,j}\det\left[c_{i,j}\right]_{i,j}\det\left[g_i(x_j)\right]_{i,j}
\]to obtain the sought formula.
\end{proof}

\begin{lemma}\label{lemmaPropKL}$\displaystyle \int_0^\infty
dx\,\KKL(x,x)=N$ and $\displaystyle \int_0^\infty
dy\,\KKL(x,y)\KKL(y,z)=\KKL(x,z)$.
\end{lemma}
\begin{proof}
From Eqs~\eqref{KLcij} and \eqref{KLcij}, we can write
\[
\int_0^\infty dx\,\KKL(x,x)=\sum_{i,j=1}^NC_{i,j}\int_0^\infty
dx\,e^{-x-a_ix}x^{\alpha+j-1}=\sum_{i,j=1}^NC_{i,j}\frac{(\alpha+j-1)!}{(1+a_i)^{\alpha+j}},
\]
which can be shown to be equal to $N$.  Similarly
\[\begin{split}
\int_0^\infty
dy\,\KKL(x,y)\KKL(y,z)&=(xz)^{\alpha/2}e^{-(x+z)/2}\sum_{i,j,k,l}e^{-a_ix}C_{i,j}C_{k,l}z^{l-1}\int_0^\infty
dy\,e^{-y-a_ky}y^{\alpha+j-1}\\
&=(xz)^{\alpha/2}e^{-(x+z)/2}\sum_{i,j,k,l}e^{-a_ix}C_{i,j}C_{k,l}z^{l-1}\frac{(\alpha+j-1)!}{(1+a_k)^{\alpha+j}}
\end{split}\]Direct manipulations then give
\[\sum_{j,k}C_{i,j}C_{k,l}\frac{(\alpha+j-1)!}{(1+a_k)^{\alpha+j}}=C_{i,l}\]
and the proof is completed.
\end{proof}
\begin{proposition}\label{CorrL}
Let $\rho^\alpha_n(x_1,\ldots,x_n)$ be the $n$ point correlation
function in the multiple   Laguerre ensemble; that is,
\[
\rho^\alpha_n(x_1,\ldots,x_n):=\frac{\MM!}{(\MM-n)!}\int_0^\infty
dx_{n+1}\cdots \int_0^\infty dx_{\MM}
P^\alpha_{\MM}(x_1,\ldots,x_{\MM}).
\]Then
\[
\rho^\alpha_n(x_1,\ldots,x_n)=\det
\left[\KKL(x_i,x_j)\right]_{i,j=1}^n.\]
\end{proposition}
\begin{proof} On the one hand,  we know from Lemma \ref{lemmadetKLP} that
\[
\rho^\alpha_n(x_1,\ldots,x_n):=\frac{1}{(\MM-n)!}\int_0^\infty
dx_{n+1}\cdots \int_0^\infty dx_{\MM}
\det\left[\KKL(x_i,x_j)\right]_{i,j=1}^{\MM}.
\]
On the other hand, Lemma \ref{lemmaPropKL} and the Laplace expansion
of the determinant imply that
\[ \int_{0}^\infty
dx_k\,\det\left[\KKL(x_i,x_j)\right]_{i,j=1}^{k}=(\MM
+1-k)\det\left[\KKL(x_i,x_j)\right]_{i,j=1}^{k-1}.\] Thus, starting
with $k=\MM$ and  applying $\MM-n-1$ times the last equation on
$\det\left[\KKL(x_i,x_j)\right]_{i,j=1}^{\MM}$ give the formula we
wanted to prove.
\end{proof}

\subsection{Multiple Hermite kernel}

The appropriate generalized  Hermite kernel has been
given in double contour integral form
by Zinn-Justin \cite{Zinn}.  With our notation and
normalization, it reads
\begin{equation}
\label{multiHK}
\KKH(x,y)=e^{y^2/2-x^2/2}\int_{\mathcal{C}_\mathbf{b}}\frac{dz}{2\pi\mathrm{i}}\int_{-\mathrm{i}\infty}^{\mathrm{i}\infty}
\frac{dw}{2\pi\mathrm{i}} \frac{e^{w^2/4-z^2/4}e^{-xz+yw}}{w-z}
\prod_{i=1}^D\left(\frac{w-b_i}{z-b_i}\right)^{m_i},\end{equation}
where it is understood that $z$ never crosses the $w$'s path.

The following proposition will considerably simplify our analysis of
the multiple Hermite kernel.  It is proved by following the method
exposed in Proposition \ref{LtoH}.
\begin{proposition}\label{KKLtoKKH}
Let $X$ be a real number such that $|\sqrt{2\alpha}X|<\alpha$.
Then, as $\alpha\rightarrow\infty$,
\[\sqrt{2\alpha}\KKL(\alpha+\sqrt{2\alpha}X,\alpha+\sqrt{2\alpha}Y)\Big|_{b_i\mapsto
b_i/\sqrt{2\alpha}}=\KKH(X,Y)+\order{\frac{1}{\sqrt{\alpha}}}.\]
\end{proposition}

\begin{corollary}  We have
\[
\KKH(x,y)=\frac{-1}{2^{\MM}\sqrt{\pi}(\MM-1)!}\frac{e^{-(x^2+y^2)/2}}{x-y}\left(
\HHt(x)
\HH(y)-\sum_{i=1}^D\frac{m_i}{\MM}\widetilde{\mathcal{H}}_{\mathbf{m}+\mathbf{e}_i}(x){\mathcal{H}}_{\mathbf{m}-\mathbf{e}_i}(y)\right)\]
and
\[\KKH(x,y)=\frac{e^{-(x^2+y^2)/2}}{2^{\MM}\sqrt{\pi}}\sum_{i=1}^D\sum_{j=1}^{m_i}
\frac{2^je^{-b_ix}}{(\MM-j)!}\,\widetilde{\mathcal{H}}_{\mathbf{m}-(j-1)\mathbf{e}_i,i}(x){\mathcal{H}}_{\mathbf{m}+j\mathbf{e}_i}(y).\]
\end{corollary}
\begin{proof}  We first make the following change of variable in Proposition
\ref{DarbouxL}: $x\mapsto \alpha+\sqrt{2\alpha}x$,
$\alpha+\sqrt{2\alpha}y$, and $b_i\mapsto b_i/\sqrt{2\alpha}$.  Then
we use the asympotics of the multiple Laguerre functions given  in
Proposition \ref{LtoH} to write the scaled perturbed Laguerre kernel
in terms of multiple Hermite functions. The proof ends  with the
comparison of the latter expression and the scaled kernel of
Proposition \ref{KKLtoKKH}.
\end{proof}

In a similar way we can prove that the correlation functions in the
perturbed Hermite ensemble are determinants of the multiple Hermite
kernel.  We recall that this ensemble is characterized by the
p.d.f.
\begin{equation}\label{pdfH}
P_N(x_1,\dots,x_N)=\frac{1}{Z_N}\prod_{i=1}^N e^{-x_i^2}\prod_{1\leq i<j\leq
N}\frac{x_j-x_i}{a_j-a_i}\det\left[e^{-a_ix_j}\right]_{i,j=1}^N.
\end{equation}
The normalization constant $Z_N$ is read off from (\ref{S1}),
\[Z_N=\frac{(-1)^{N(N-1)/2} (2\pi)^{N/2}N!}{2^{N^2/2}}\prod_{i=1}^N
e^{a_i^2/4}.\]
\begin{corollary}Let
\[
\rho_n(x_1,\ldots,x_n):=\frac{\MM!}{(\MM-n)!}\int_{-\infty}^\infty
dx_{n+1}\cdots \int_{-\infty}^\infty dx_{\MM}
P_{\MM}(x_1,\ldots,x_{\MM}).
\]Then
\[
\rho_n(x_1,\ldots,x_n)=\det \left[\KKH(x_i,x_j)\right]_{i,j=1}^n.\]
\end{corollary}
\begin{proof}The expressions of the Laguerre and Hermite
p.d.f.~(respectively given in Eqs \eqref{pdfL} and \eqref{pdfH})
together with the Stirling approximation  allow us to write
\[\lim_{\alpha\rightarrow\infty}
\left\{(2\alpha)^{N/2}P^\alpha_N(x_1,\ldots,x_N)\Big|_{\substack{x_i\mapsto\alpha+\sqrt{2\alpha}x_i\\b_i\mapsto
b_i/\sqrt{2\alpha}}}\right\} =P_N(x_1,\ldots,x_N).\]Hence the
corollary is an immediate consequence of Propositions \ref{CorrL} and
\ref{KKLtoKKH}.
\end{proof}

\section{Quasi-null case}

In this section, we suppose  that  all but a fixed number
of eigenvalues of the
matrix  $\mathbf{A}$ are zero.  Precisely,
\begin{equation}\label{QNC}\begin{split}
\aaa\,&=(a_1,\ldots,a_r,0,\ldots,0),\\
\bb\,&=(b_1,\ldots,b_d,0),\\
\mm&=(m_1,\ldots,m_d,N-r).
\end{split}\end{equation}
Here $d=D-1$ gives the number of distinct parameters in $\mathbf{a}$ that are not equal to zero. To make
contact with the results of \cite{Baik2004} and \cite{Peche}
reviewed in \S 1, we seek limiting expressions for the kernels, and
thus for the correlations, when the size $N$ of the random matrices
becomes infinite.

\subsection{Perturbed Laguerre ensemble}

The kernel is now given by
\begin{equation}\label{QNLK}
\KKL(x,y)=\sqrt{\frac{w_\alpha(x)}{w_\alpha(y)}}
\int_{\mathcal{C}^{\{-1\}}_{\mathbf{a}}}\frac{dz}{2\pi\mathrm{i}}\int_{\mathcal{C}_{\{-1\}}}\frac{dw}{2\pi\mathrm{i}}
\frac{e^{-xz+yw}}{w-z}\left(\frac{1+z}{1+w}\right)^{N+\alpha}\left(\frac{w}{z}\right)^{N-r}
\prod_{i=1}^r\frac{w-a_i}{z-a_i},
\end{equation}
where $N=\MM$ and $w_\alpha(x)=x^\alpha e^{-x}$, as usual. It is
convenient to introduce
\begin{equation}\label{bKKL}
{\bKKL}(x,y):=\sqrt{\frac{w_\alpha(y)}{w_\alpha(x)}}\KKL(x,y).
\end{equation}
This gauge transformation does not affect the correlation functions
because
\[\det\left[\frac{f(x_i)}{f(x_j)}g(x_i,y_j)\right]_{i,j=1}^N=
\det\left[g(x_i,y_j)\right]_{i,j=1}^N\]
for all $g(x_i,y_j)$ and $f(x)\neq 0$.

A striking feature of the quasi-null soft edge scaled correlation
(\ref{3.a}) is that the correlation kernel consists of the null case
Airy kernel (\ref{3.a2}) plus $r$ perturbation terms. As our key
result we now show that at the level of finite $N$, the multiple
Laguerre kernel can be decomposed as a (null case) Laguerre kernel
plus $r$ perturbation terms.  The latter are written in terms of
incomplete multiple Laguerre functions of type I and II,
\begin{equation}\label{tLambdai}
\widetilde{\Lambda}^{(i)}(x):=\int_{\mathcal{C}_{\{0,a_1,\ldots,a_i\}}}\frac{dz}{2\pi\mathrm{i}}\frac{e^{-xz}(1+z)^{N+\alpha}}{z^{N-r}\prod_{k=1}^i(z-a_k)}
\end{equation}
and
\begin{equation}\label{Lambdai}
{\Lambda}^{(i)}(x):=\int_{\mathcal{C}_{\{-1\}}}\frac{dw}{2\pi\mathrm{i}}\frac{e^{xw}w^{N-r}\prod_{k=1}^{i-1}(w-a_k)}{(1+w)^{N+\alpha}}
\end{equation}
respectively. Comparison with (\ref{multiLaguerreI}) shows that
\begin{equation}\label{as.1}
\tilde{\Lambda}^{(i)}(x) = \tilde{\mathcal L}_{\mathbf{m}'}^{\alpha +
r +1-i}(x|\mathbf{b}') \end{equation} 
when
\begin{equation}\label{bprime}
\mathbf{b}'=\{a_1,\ldots,a_i,0\} 
\quad\mbox{and}\quad \mathbf{m}'=\{\overbrace{1,\ldots,1}^i,N-r\}.
\end{equation}
Similarly, we see from (\ref{multiLaguerreII}) that
\begin{equation}\label{as.2}
{\Lambda}^{(i+1)}(x) = \frac{(N+i-r)!}{(N+\alpha - 1)!} \, x^{\alpha + r - i-1} e^{-x}\,
{\mathcal L}_{\mathbf{m}'}^{\alpha + r - i-1 }(x|\mathbf{b}').\end{equation}
As $ \tilde{\mathcal L}_{\mathbf m}^\alpha(x)$ and ${\mathcal
L}_{\mathbf m}^\alpha(x)$ can be interpreted as anti-derivatives and
derivations of Laguerre polynomials according to (\ref{3.12}) and
(\ref{3.13}), it follows from (\ref{as.1}) and (\ref{as.2}) that
$\widetilde{\Lambda}^{(i)}(x)$ and $\Lambda^{(i)}(x)$ are also simply related to
 Laguerre polynomials. Explicitly,
\[\mathcal{D}_{-a_1,x}\cdots\mathcal{D}_{-a_i,x}\left[\widetilde{\Lambda}^{(i)}(x)\right]=L^{\alpha+r+1}_{N-r-1}(x)\]
and
\[{\Lambda}^{(i)}(x)=(-1)^{\alpha+r+i}e^{-x}\mathcal{D}_{a_1,x}\cdots\mathcal{D}_{a_{i-1},x}\left[L^{-\alpha-r+1}_{N-\alpha+1}(x)\right].\]
where in deriving the latter use has also been made of (\ref{3.8a}).

\begin{proposition}\label{propKLambda}
Let $\bar{K}^{\alpha}_N(x,y)$ be the gauged transformed Laguerre
kernel;  that is,
\[
\bar{K}^{\alpha}_N(x,y)
=\lim_{\mathbf{a}\rightarrow\mathbf{0}}\bKKL(x,y)
=\int_{\mathcal{C}^{\{-1\}}_{\{0\}}}\frac{dz}{2\pi\mathrm{i}}\int_{\mathcal{C}_{\{-1\}}}\frac{dw}{2\pi\mathrm{i}}
\frac{e^{-xz+yw}}{w-z}\left(\frac{1+z}{1+w}\right)^{N+\alpha}\left(\frac{w}{z}\right)^{N}\]for
$N=\MM$ and
$\mathcal{C}^{\{-1\}}_{\{0\}}\cap\mathcal{C}_{\{-1\}}=\emptyset$.
Then
\begin{equation}\label{5.4'}\bKKL(x,y)=\bar{K}^{\alpha+r}_{N-r}(x,y)+\sum_{i=1}^r\widetilde{\Lambda}^{(i)}(x){\Lambda}^{(i)}(y).
\end{equation}
\end{proposition}
\begin{proof}
We first consider the relation
\begin{equation}\label{eqProdSum}
\frac{1}{w-z}\prod_{i=1}^r\frac{w-a_i}{z-a_i}=\frac{1}{w-z}+\sum_{i=1}^r\frac{\prod_{k=1}^{i-1}w-a_k}{\prod_{k=1}^{i}z-a_k}.\end{equation}For
$r=1$, this reads
\[\frac{1}{w-z}\frac{w-a_1}{z-a_1}=\frac{1}{w-z}+\frac{1}{z-a_1},\]which is
trivially true.  For $r>1$, Eq.~\eqref{eqProdSum} is proved by
induction; explicitly,
\[\begin{split}
\frac{1}{w-z}\prod_{i=1}^{r+1}\frac{w-a_i}{z-a_i}&=\frac{1}{w-z}\left(\prod_{i=1}^{r}\frac{w-a_i}{z-a_i}\right)\frac{z-a_{r+1}+w-z}{z-a_{r+1}}\\
&=\frac{1}{w-z}\prod_{i=1}^{r}\frac{w-a_i}{z-a_i}+\left(\prod_{i=1}^{r}\frac{w-a_i}{z-a_i}\right)\frac{1}{z-a_{r+1}}\\
&=\frac{1}{w-z}+\sum_{i=1}^r\frac{\prod_{k=1}^{i-1}w-a_k}{\prod_{k=1}^{i}z-a_k}+\frac{\prod_{k=1}^{r}w-a_k}{\prod_{k=1}^{r+1}z-a_k}
\end{split}\]as expected.

We then substitute Eq.~\eqref{eqProdSum} in the quasi-null kernel,
given in Eq.~\eqref{QNLK} (see also Eq.~\eqref{bKKL}).  This gives
\begin{multline*}\bKKL(x,y)=
\int_{\mathcal{C}^{\{-1\}}_{\{0\}}}\frac{dz}{2\pi\mathrm{i}}\int_{\mathcal{C}_{\{-1\}}}\frac{dw}{2\pi\mathrm{i}}
\frac{e^{-xz+yw}}{w-z}\left(\frac{1+z}{1+w}\right)^{N+\alpha}\left(\frac{w}{z}\right)^{N-r}\\
+\sum_{i=1}^r\int_{\mathcal{C}_{\{a_1,\ldots,a_i,0\}}}\frac{dz}{2\pi\mathrm{i}}\frac{e^{-xz}(1+z)^{N+\alpha}}{z^{N-r}\prod_{k=1}^{i}(z-a_k)}
\int_{\mathcal{C}_{\{-1\}}}\frac{dw}{2\pi\mathrm{i}}\frac{e^{yw}w^{N-r}\prod_{k=1}^{i-1}w-a_k}{(1+w)^{N+\alpha}}.\end{multline*}
Comparing the latter equation with Eqs \eqref{tLambdai},
\eqref{Lambdai} finally  completes the proof.
\end{proof}

\subsection{Perturbed Hermite kernel}
When the set $\aaa$ of eigenvalues is quasi-null, the multiple
Hermite kernel becomes
\begin{equation}
\label{QNHK}
\KKH(x,y)=e^{y^2/2-x^2/2}\int_{\mathcal{C}_\mathbf{a}}\frac{dz}{2\pi\mathrm{i}}\int_{-\mathrm{i}\infty}^{\mathrm{i}\infty}
\frac{dw}{2\pi\mathrm{i}}
\frac{e^{w^2/4-z^2/4}e^{-xz+yw}}{w-z}\left(\frac{w}{z}\right)^{N-r}
\prod_{i=1}^r\left(\frac{w-a_i}{z-a_i}\right).\end{equation} Note
that the $w$'s contour integral does note pass through the origin.
As in the perturbed Laguerre ensemble, we introduce incomplete
multiple Hermite functions,
\begin{equation}\label{tGAMMAi}
\widetilde{\Gamma}^{(i)}(x):=e^{-x^2/2}\int_{\mathcal{C}_{\{0,a_1,\ldots,a_i\}}}\frac{dz}{2\pi\mathrm{i}}\frac{e^{-xz-z^2/4}}{z^{N-r}\prod_{k=1}^i(z-a_k)}
\end{equation}
and
\begin{equation}\label{GAMMAi}
{\Gamma}^{(i)}(x):=e^{x^2/2}\int_{-\mathrm{i}\infty}^{\mathrm{i}\infty}
\frac{dw}{2\pi\mathrm{i}}{e^{xw+w^2/4}w^{N-r}\prod_{k=1}^{i-1}(w-a_k)}.
\end{equation}
Comparison with (\ref{multiHermiteI}) and (\ref{multiHermiteII})
shows
\begin{equation}
\begin{split}
&\tilde{\Gamma}^{(i)}(x)  =  \frac{e^{-x^2/2} }{ (-2)^{N+i-r-1}
(N+i-r-1)!} \tilde{\mathcal H}_{\mathbf{ m}'}(x|\mathbf{b}') , \\
&{\Gamma}^{(i+1)}(x)  = \frac{(-1)^{N+i-r} }{ \sqrt{\pi} }
e^{-x^2/2} {\mathcal H}_{\mathbf{m}'}(x|\mathbf{b}'),
\end{split}
\end{equation}
where $\mathbf{b}'$ and $\mathbf{m}'$ are the restricted sets of parameters given in \eqref{bprime}.
Consequently, we read off from (\ref{m.1}), (\ref{m.2}) that these
special multiple Hermite polynomials are related to the classical
Hermite polynomials by
\[H_{N-r-1}=(-2)^{N-r-1}(N-r-1)!\mathcal{D}_{-a_1,x}\cdots\mathcal{D}_{-a_i,x}\left[e^{x^2/2}\widetilde{\Gamma}^{(i)}(x)\right]\]
and
\[{\Gamma}^{(i)}(x)=\frac{(-1)^{N+r+i+1}}{\sqrt{\pi}}e^{x^2/2}\mathcal{D}_{a_1,x}\cdots\mathcal{D}_{a_{i-1},x}\left[e^{-x^2/2}H_{N-r}(x)\right].\]

The algebraic property given in Eq.\ \eqref{eqProdSum} can be used
to prove the following proposition.

\begin{proposition}\label{PropHpertubed}
Let ${K}_N(x,y)$ be the usual Hermite kernel; that is,
\[{K}_N(x,y)=\lim_{\mathbf{a}\rightarrow\mathbf{0}}\KKH(x,y)=e^{y^2/2-x^2/2}\int_{\mathcal{C}_{\{0\}}}\frac{dz}{2\pi\mathrm{i}}\int_{-\mathrm{i}\infty}^{\mathrm{i}\infty}
\frac{dw}{2\pi\mathrm{i}}
\frac{e^{w^2/4-z^2/4}e^{-xz+yw}}{w-z}\left(\frac{w}{z}\right)^{N}.\]
Then
\[\KKH(x,y)={K}_{N-r}(x,y)+\sum_{i=1}^r\widetilde{\Gamma}^{(i)}(x){\Gamma}^{(i)}(y).\]
\end{proposition}

Before considering the large $N$ limit of the  kernels, let us
link  the perturbed Laguerre and Hermite ensembles by relating the incomplete multiple
Laguerre and Hermite polynomials.

\begin{proposition}Set $X$ a real number satisfying
$|\sqrt{2\alpha}X|<\alpha$.  Then, as $\alpha\rightarrow\infty$,
\[(2\alpha)^{(1+r-i-N)/2}\,\widetilde{\Lambda}^{(i)}(\alpha+\sqrt{2\alpha}X)\Big|_{a_i\mapsto
a_i/\sqrt{2\alpha}}=e^{X^2/2}\,\widetilde{\Gamma}^{(i)}(X)+\order{\frac{1}{\sqrt{\alpha}}}\]and
\[(2\alpha)^{(N+i-r)/2}\,{\Lambda}^{(i)}(\alpha+\sqrt{2\alpha}Y)\Big|_{a_i\mapsto
a_i/\sqrt{2\alpha}}=e^{-Y^2/2}\,{\Gamma}^{(i)}(Y)+\order{\frac{1}{\sqrt{\alpha}}}.\]
\end{proposition}
\begin{proof}We essentially follow the steps given in the proof of
Proposition
\ref{LtoH}.  We skip the detail. \end{proof}

The latter result obviously complies with Proposition \ref{KKLtoKKH}
and the  asymptotics
\[\sqrt{\frac{w_\alpha(\alpha+\sqrt{2\alpha}X)}{w_\alpha(\alpha+\sqrt{2\alpha}Y)}}=e^{(Y^2-X^2)/2}+\order{\frac{1}{\sqrt{\alpha}}}.\]

\section{Kernels at the edges of the spectrum}

\subsection{Hard edge}

We know from (\ref{3.a2}) that the null-case hard edge kernel can be
written in terms of Bessel functions $J_\alpha(x)$. The latter satisfy
\begin{equation}\label{Bessel}
J_\alpha(x)=\int_{\mathcal{C}_{\{0\}}}\frac{dz}{2\pi\mathrm{i}}\frac{e^{x(z-z^{-1})/2}}{z^{\alpha+1}},\quad
J_{-\alpha}(x)=(-1)^{\alpha}J_\alpha(x),\quad\alpha\in\mathbb{Z}.
\end{equation}
The form (\ref{5.4'}) of the perturbed Laguerre kernel suggests that
its hard edge scaling can be expressed in terms of the Bessel like
functions (\ref{3.b3}), (\ref{3.b4}). Use of (\ref{3.9'}) and
(\ref{Bessel}) shows these functions are simply related to the
Bessel function by
\[
J_{\alpha+r+1}(\sqrt{x})=(4x)^{(\alpha+r+1)/2}
\mathcal{D}_{-h_1,x}\cdots\mathcal{D}_{-h_i,x}\left[\widetilde{J}^{(i)}(x)\right]\]
and
\[{J}^{(i)}(x)=(-1)^{i-1}\mathcal{D}_{h_1,x}\cdots\mathcal{D}_{h_{i-1},x}\left[x^{(\alpha+r-1)/2}J_{\alpha+r-1}(\sqrt{x})\right].\]

The functions $\widetilde{J}^{(i)}$ and $J^{(i)}$ are in fact
incomplete versions of the multiple Bessel functions of type I and
II, which we respectively define as
\begin{equation}\label{BesselI}
\widetilde{\mathcal{J}}^\alpha_\mm(x)=\widetilde{\mathcal{J}}^{\alpha}_\mm(x|\boldsymbol{\nu})
:=(2x)^{\alpha}\int_{\mathcal{C}_{\{0\}\cup\boldsymbol{\nu}}}\frac{dz}{2\pi\mathrm{i}}\frac{z^{\MM+\alpha-1}e^{-x^2z+1/(4z)}}{\prod_{i=1}^D(z-\nu_i)^{m_i}}
\end{equation}
and
\begin{equation}\label{BesselII}
{\mathcal{J}}^{\alpha}_\mm(x)={\mathcal{J}}^{\alpha}_\mm(x|\boldsymbol{\nu})
:=(2x)^{-\alpha}\int_{\mathcal{C}_{\{0\}}}\frac{dw}{2\pi\mathrm{i}}\frac{e^{x^2w-1/(4w)}\prod_{i=1}^D(w-\nu_i)^{m_i}}{w^{\MM+\alpha+1}}.
\end{equation}
Recall that $D$ denotes, as in Eqs.\ \eqref{aset} and \eqref{bset}, the number of distinct parameters in $\mathbf{h}={\boldsymbol{\nu}}^{\mathbf{m}}$.   The link relation between the incomplete and the complete functions
is obvious,
\begin{equation}\label{BesselIncomplete}
\widetilde{\mathcal{J}}^{\alpha+r+1-i}_{\{1^i\}}(\sqrt{x})=(2\sqrt{x})^{\alpha+r+1-i}\widetilde{J}^{(i)}(x)\quad\text{and}\quad
 {\mathcal{J}}^{\alpha+r-i}_{\{1^{i-1}\}}(\sqrt{x})=(2\sqrt{x})^{-\alpha-r+i}{J}^{(i)}(x)
,\end{equation}where $\{1^{k}\}$ denotes a composition of length $k$
in which  all entries being equal to unity.

\begin{lemma}\label{lemmaHard}Suppose that Eq.~\eqref{QNC} holds.  Let $b_i=4N\nu_i$ for $i=1,\ldots,d$ and $\boldsymbol{\mu}=(m_1,\ldots,m_d)$.  Then, as
$N\rightarrow\infty$,
\[\begin{split}
(4N)^{-\alpha}\widetilde{\mathcal{L}}^\alpha_\mm(X/4N)&=(2\sqrt{X})^{-\alpha}\widetilde{\mathcal{J}}^\alpha_{\boldsymbol{\mu}}(\sqrt{X})+\order{1/{N}},\\
(4N)^{-\alpha}{\mathcal{L}}^\alpha_\mm(X/4N)&=(2\sqrt{X})^{-\alpha}{\mathcal{J}}^\alpha_{\boldsymbol{\mu}}(\sqrt{X})+\order{1/{N}}.\end{split}\]
Similarly,  for $a_i=4Nh_i$ ($i=1,\ldots,r$), we have
\[\begin{split}
(4N)^{i-\alpha-r-1}\widetilde{\Lambda}^{(i)}(X/4N)&=\widetilde{J}^{(i)}(X)+\order{1/{N}},\\
(4N)^{\alpha+r-i} {\Lambda}^{(i)}(X/4N)&={J}^{(i)}(X)+\order{1/{N}}.
\end{split}\]
\end{lemma}
\begin{proof}From  the definition of $\widetilde{\mathcal{L}}^\alpha_\mm$, given
in Eq~\eqref{multiLaguerreI}, we have
\[\widetilde{\mathcal{L}}^\alpha_\mm(X/4N)
=\int_{\mathcal{C}^{\{-1\}}_{\{0,b_1,\ldots,b_d\}}}\frac{dz}{2\pi\mathrm{i}}\frac{e^{-Xz/4N}(1+z)^{N+\alpha-1}}{z^{N-r}\prod_{k=1}^d(z-b_k)^{m_k}}.\]
Then, changing $z$ for $4Nw$ and $b_i$ for $4N\nu_i$, we get
\[\widetilde{\mathcal{L}}^\alpha_\mm(X/4N)=
(4N)^{\alpha}\int_{\mathcal{C}_{\{0,h_1,\ldots,h_d\}}}\frac{dw}{2\pi\mathrm{i}}\frac{e^{-Xw}w^{\alpha+r-1}(1+1/4Nw)^{N+\alpha}}{\prod_{k=1}^d(w-\nu_k)^{m_k}}.\]
But for $N\rightarrow\infty$,
\[
(1+1/4Nw)^{N+\alpha}=(1+1/4Nw)^{N}+\order{1/N}
=e^{1/4w}+\order{1/N}.
\]
This immediately implies that
\[\widetilde{\mathcal{L}}^\alpha_\mm(X/4N)=
(4N)^{\alpha}\int_{\mathcal{C}_{\{0,h_1,\ldots,h_d\}}}\frac{dw}{2\pi\mathrm{i}}\frac{e^{-Xw+1/4w}w^{\alpha+r-1}}{\prod_{k=1}^d(w-\nu_k)^{m_k}}+\order{N^{\alpha-1}}.\]
Comparing with Eq.~\eqref{BesselII}  gives the desired result. The
proof for the other functions is similar.
\end{proof}

\begin{theorem}\label{TheoHard}Suppose that Eq.~\eqref{QNC} holds.  Set $a_i=4Nh_i$ for $i=1,\ldots,q$ and  $a_i=(4N)^2h_i$ for $i=q+1,\ldots,r$.
Following the notation of \S 3, let
$\mathbf{h}:=\{h_1,\ldots,h_q\}=\{\nu_i^{\mu_i},\ldots,\nu_p^{\mu_q}\}=:\boldsymbol{\nu}^{\boldsymbol{\mu}}$
 with $q=|\boldsymbol{\mu}|$.
Then, as $N\rightarrow\infty$,
\[\frac{1}{4N}\bKKL\left(\frac{X}{4N},\frac{Y}{4N}\right)
=\mathcal{K}^{\mathrm{hard}}_{\boldsymbol{\mu}}(X,Y)+\order{\frac{1}{N}},\]
where
\begin{align}\mathcal{K}^{\mathrm{hard}}_{\boldsymbol{\mu}}(X,Y)&
=\left(\frac{Y}{X}\right)^{(\alpha+r)/2}K^{\mathrm{hard}}_{\alpha+r}(X,Y)+\sum_{i=1}^{q}\widetilde{J}^{(i)}(X){J}^{(i)}(Y)\label{Hard1st}\\
&=\int_{\mathcal{C}_{\{0\}\cup\boldsymbol{\nu}}}\frac{dz}{2\pi\mathrm{i}}\int_{\mathcal{C}^{\{z\}}_{\{0\}}}\frac{dw}{2\pi\mathrm{i}}
\frac{e^{-Xz+1/(4z)}e^{Yw-1/(4w)}}{w-z}\left(\frac{z}{w}\right)^{\alpha+r}\prod_{i=1}^p\left(\frac{w-\nu_i}{z-\nu_i}\right)^{\mu_i}\label{Hard2nd}\\
&= \frac{1}{2(X-Y)}\left(\frac{Y}{X}\right)^{\alpha'/2}\left(J(X,Y)
-2\sum_{i=1}^{q}\mu_i\widetilde{\mathcal{J}}^{\alpha'}_{\boldsymbol{\mu}+\mathbf{e}_i}(\sqrt{X}){\mathcal{J}}^{\alpha'}_{\boldsymbol{\mu}-\mathbf{e}_i}(\sqrt{Y})\right).
\label{Hard3rd}\end{align}with $\alpha'=\alpha+r-q$ and
\begin{multline}J(X,Y)=\widetilde{\mathcal{J}}^{\alpha'}_{\boldsymbol{\mu}}(\sqrt{X})\left((\alpha'+q){\mathcal{J}}^{\alpha'}_{\boldsymbol{\mu}}(\sqrt{Y})
-\sqrt{Y}{\mathcal{J}}^{\alpha'+1}_{\boldsymbol{\mu}}(\sqrt{Y})\right)\\
+
{\mathcal{J}}^{\alpha'}_{\boldsymbol{\mu}}(\sqrt{Y})\left((\alpha'+q)\widetilde{\mathcal{J}}^\alpha_{\boldsymbol{\mu}}(\sqrt{X})
-\sqrt{X}\widetilde{\mathcal{J}}^{\alpha'-1}_{\boldsymbol{\mu}}(\sqrt{X})\right).\end{multline}
\end{theorem}
\begin{proof}
Let us begin with  Eq.~\eqref{Hard1st}. We know from the previous
lemma that
\[
\widetilde{\Lambda}^{(i)}(X/4N){\Lambda}^{(i)}(Y/4N)=(4N)\widetilde{J}^{(i)}(X){J}^{(i)}(Y)+\order{1}
\quad\mbox{when}\quad i=1,\ldots,q.  \] For $i>q$, Lemma
\ref{LemmaSoft} and Eqs  (\ref{3.b2})--(\ref{3.b3}) imply
\begin{multline*}
\widetilde{\Lambda}^{(i+q)}(X/4N){\Lambda}^{(i+q)}(Y/4N)=
\widetilde{J}^{(q)}(X){J}^{(q)}(Y)+\order{\frac{1}{N}}
\quad\mbox{when}\quad i=1,\ldots,r-q.  \end{multline*}Furthermore,
we have from Eq. \eqref{bKKL} and \S 2 (see also
\cite[\S4]{ForresterBook})
\[\frac{1}{4N}\bar{K}_N^{\alpha+r}
\left(\frac{X}{4N},\frac{Y}{4N}\right)=\left(\frac{X}{Y}\right)^{(\alpha+r)/2}K^{\mathrm{hard}}_{\alpha+r}(X,Y)+\order{\frac{1}{N}},\]where
the hard edge kernel is defined in Eq.~\eqref{3.b2}. We finally
obtain  Eq.~\eqref{Hard1st} by using Proposition \ref{propKLambda}.

We now turn our attention to Eq.~\eqref{Hard2nd}.  We first claim
that the hard edge kernel has the  double integral
representation
\[K^{\mathrm{hard}}_{\alpha}(x,y)=\left(\frac{x}{y}\right)^{\alpha/2}\int_{\mathcal{C}_{\{0\}}}\frac{dz}{2\pi\mathrm{i}}\int_{\mathcal{C}^{\{z\}}_{\{0\}}}\frac{dw}{2\pi\mathrm{i}}
\frac{e^{-xz+1/(4z)}e^{yw-1/(4w)}}{w-z}\left(\frac{z}{w}\right)^{\alpha}.\]
This is proved by effectively multiplying the previous equation by $(x-y)/(x-y)$
and integrating by parts; that is, {\small\begin{align}
K^{\mathrm{hard}}_{\alpha}(x,y)&=\frac{1}{x-y}\left(\frac{x}{y}\right)^{\alpha/2}\int_{\mathcal{C}_{\{0\}}}\frac{dz}{2\pi\mathrm{i}}\int_{\mathcal{C}^{\{z\}}_{\{0\}}}\frac{dw}{2\pi\mathrm{i}}
\frac{e^{1/(4z)-1/(4w)}}{w-z}\left(\frac{z}{w}\right)^{\alpha}\left(-\frac{\partial}{\partial
z}-\frac{\partial}{\partial w}\right)\left(e^{-xz+yw}\right),\nonumber\\
&=\frac{1}{x-y}\left(\frac{x}{y}\right)^{\alpha/2}\int_{\mathcal{C}_{\{0\}}}\frac{dz}{2\pi\mathrm{i}}\int_{\mathcal{C}^{\{z\}}_{\{0\}}}\frac{dw}{2\pi\mathrm{i}}
{e^{1/(4z)-1/(4w)}}e^{-xz+yw}\left(\frac{z}{w}\right)^{\alpha}\left(\frac{\alpha}{zw}-\frac{z+w}{4z^2w^2}\right),\nonumber\\
&=\frac{1}{2(x-y)}\left(2\alpha
J_\alpha(\sqrt{x})J_\alpha(\sqrt{y})-\sqrt{x}J_{\alpha-1}(\sqrt{x})J_\alpha(\sqrt{y})-J_\alpha(\sqrt{x})\sqrt{y}J_{\alpha+1}(\sqrt{y})\right),\nonumber\end{align}}
which, by virtue of
$\sqrt{x}J_\alpha'(\sqrt{x})=\sqrt{x}J_{\alpha-1}(\sqrt{x})-\alpha
J_\alpha(\sqrt{x})$ and
$\sqrt{x}J_{\alpha+1}(\sqrt{x})=-\sqrt{x}J_{\alpha-1}(\sqrt{x})+2\alpha
J_\alpha(\sqrt{x})$, turns out to be equivalent to Eq.~\eqref{3.b2}.
Then we use the integral representations of the incomplete multiple
Bessel functions, which are given in Eqs \eqref{3.b3}--\eqref{3.b4},
to rewrite Eq.~\eqref{Hard1st} as
\begin{multline*}
\mathcal{K}^{\mathrm{hard}}_{\alpha}(X,Y)=\int_{\mathcal{C}_{\{0\}\cup\boldsymbol{h}}}\frac{dz}{2\pi\mathrm{i}}\int_{\mathcal{C}^{\{z\}}_{\{0\}}}\frac{dw}{2\pi\mathrm{i}}
{e^{-Xz+1/(4z)}e^{Yw-1/(4w)}}\\
\times\left(\frac{z}{w}\right)^{\alpha+r}\left(\frac{1}{w-z}+\sum_{i=1}^q\prod_{k=1}^i\left(\frac{w-h_k}{z-h_k}\right)\frac{1}{w-h_i}\right),
\end{multline*}
where the Cauchy theorem has been applied for deforming the
 contours
 $\mathcal{C}_{\{0\}}$ and $\mathcal{C}_{\{0,h_1,\ldots,h_i\}}$
into $\mathcal{C}_{\{0\}\cup\boldsymbol{h}}$.  Use of
Eq.~\eqref{eqProdSum} then finishes the proof of formula
Eq.~\eqref{Hard2nd}.

We show Eq. \eqref{Hard3rd} with the same method than the one
explained in the proof of Proposition \ref{DarbouxL}: we multiply
Eq.~\eqref{Hard2nd} by
$(X-Y)^{-1}(X-Y)=-(X-Y)^{-1}e^{Xz-Yw}(\partial_z+\partial_w)e^{-Xz+Yw}$;
we integrate by parts; and we compare the result with the definition of
the multiple Bessel functions, Eqs
\eqref{BesselI}--\eqref{BesselII}.
\end{proof}

\subsection{Soft edge}
Working directly with the double contour form (\ref{QNLK}) of the
correlation kernel Baik, Ben Arous and P\'ech\'e \cite{Baik2004}
have shown that the soft edge scaling (\ref{3.0}) of perturbed
Laguerre kernel can be written in terms of multiple Airy functions
according to the results (\ref{3.0})--(\ref{3.a4}). From the contour
integral representation of the Airy function,
\begin{equation}\label{Airy}
{\mathrm{Ai}}(x):=\int_{\mathcal{A}}\frac{dv}{2\pi\mathrm{i}}{e^{-xv+v^3/3}},
\end{equation}
 and (\ref{3.9'}) we
see that the multiple functions (\ref{3.a2}), (\ref{3.a3}) are
related to the Airy function by
\[\mathrm{Ai}(x)=\mathcal{D}_{-s_1}\cdots\mathcal{D}_{-s_i}\left[\widetilde{\mathrm{Ai}}^{(i)}(x)\right]
\]
and
\[
{\mathrm{Ai}}^{(i)}(x)=(-1)^i\mathcal{D}_{s_1}\cdots\mathcal{D}_{s_{i-1}}\left[\mathrm{Ai}(x)\right].\]

According to \S 6.1, the functions $\mathrm{Ai}^{(i)}$ and
$\widetilde{\mathrm{Ai}}^{(i)}$ can be completed in order to get the
multiple Airy functions of types I and II,
\begin{equation}\label{AiryI}
\Ait{\mm}(x)=\Ait{\mm}(x|\boldsymbol{\sigma}):=\int_{\mathcal{A}_{\boldsymbol{\sigma}}}\frac{dv}{2\pi\mathrm{i}}{e^{-xv+v^3/3}}{\prod_{k=1}^D(v-\sigma_k)^{-m_k}},
\end{equation}
and\begin{equation}\label{AiryII}
\Ai{\mm}(x)=\Ai{\mm}(x|\boldsymbol{\sigma}):=(-1)^{\MM}\int_{\mathcal{A}}\frac{dv}{2\pi\mathrm{i}}{e^{-xv+v^3/3}}{\prod_{k=1}^{D}(v+\sigma_k)^{m_k}}.
\end{equation}
One can easily relate the two sets of multiple Airy functions,
follows: \begin{equation}\label{AiryIncomplete}
\Ait{\{1^{i}\}}(x)=\widetilde{\mathrm{Ai}}^{(i)}(x)\quad
\text{and}\quad\Ai{\{1^{i-1}\}}(x)=- {\mathrm{Ai}}^{(i)}(x).
\end{equation}

\begin{lemma}\label{LemmaSoft}Suppose that Eq.~\eqref{QNC} holds.  Let $A=4N$ and $B=2(2N)^{1/3}$.  Let also $b_i=-1/2+\sigma_i/B$ for $i=1,\ldots,d$, $\boldsymbol{\mu}=(m_1,\ldots,m_d)$ and $|\boldsymbol{\mu}|=r$. Then, as
$N\rightarrow\infty$,
\[\begin{split}
B^{1-r}\widetilde{\mathcal{L}}^\alpha_\mm(A+BX)&=\frac{(-1)^{N+r+1}}{2^{\alpha+r-1}}\Ait{\boldsymbol{\mu}}(X)+\order{\frac{1}{N^{1/3}}}\\
B^{1+r}{\mathcal{L}}^\alpha_\mm(A+BX)&=\frac{(-1)^{N+r}}{2^{\alpha-r-1}}\Ai{\boldsymbol{\mu}}(X)+\order{\frac{1}{N^{1/3}}}.
\end{split}\]
Similarly, when $a_i=-1/2+s_i/B$ for $i=1,\ldots,r$, we have
\[\begin{split}
B^{1-i}e^{-(A+BX)/2}\widetilde{\Lambda}^{(i)}(A+BX)={(-1)^{N+r}}{2^{-\alpha-r}}\widetilde{\mathrm{Ai}}^{(i)}(X)+\order{\frac{1}{N^{1/3}}},\\
B^{i}e^{(A+BY)/2}{\Lambda}^{(i)}(A+BY)=(-1)^{N+r}{2^{\alpha+r}}{\mathrm{Ai}}^{(i)}(Y)+\order{\frac{1}{N^{1/3}}}.
\end{split}\]
\end{lemma}
\begin{proof}
The proof is based on the saddle point method.  For instance, consider
\[\widetilde{\Lambda}^{(i)}(A+BX):=\int_{\mathcal{C}_{\{0,a_1,\ldots,a_r\}}}\frac{dz}{2\pi\mathrm{i}}e^{Nf(z)}g(z)\]
with
\[f(z)=-4z+\ln(1+z)-\ln z, \quad
g(z)=\frac{e^{-BXz}z^r(1+z)^{\alpha}}{\prod_{k=1}^i(z-a_k)}.\]  The
function $f$ has a simple saddle point of order two at $z_0=-1/2$;
that is,
\[ f'(z_0)=0=f''(z_0) \quad f'''(z_0)\neq 0.\] Since $f'''(z_0)=32$,
we have that $(z-z_0)f'''(z_0)$ is minimum when
$\mathrm{arg}(z-z_0)=-\pi/3,\pi/3,\pi$.  We choose $\pm \pi/3$.
Doing the change of variable $v=B(z-z_0)$, we understand that the
steepest descent contour in the complex $v$-plane starts at $\infty
e^{-\mathrm{i}\pi/3}$ crosses the origin and ends at $\infty
e^{\mathrm{i}\pi/3}$.  Moreover,
\[Nf(z)=2N-\mathrm{i}\pi N+v^3/3+\order{N^{-1/3}},\]while
\[
g(z)=(-1)^r2^{-r-\alpha}B^{i}e^{-Xv+BX/2}\prod_{k=1}^i
(v-s_k)^{-1}\left(1+\order{N^{-1/3}}\right)\]if $s_k=B(a_k+1/2)$.
Therefore,
\begin{multline*}\widetilde{\Lambda}^{(i)}(A+BX)=(-1)^{r+N}2^{-r-\alpha}B^{i}e^{(A+BX)/2}\\
\times\int_{\mathcal{A}_{\{s_1,\ldots,s_i\}}}\frac{dv}{2\pi\mathrm{i}}{e^{-xv+v^3/3}}{\prod_{k=1}^i(v-s_k)^{-1}}\left(1+\order{\frac{1}{N^{1/3}}}\right).\end{multline*}
This proves the asymptotics of $\widetilde{\Lambda}^{(i)}$ .  The
asymptotic behavior of the other functions is proved similarly.
\end{proof}

\begin{theorem}\label{TheoSoft}Suppose that Eq.~\eqref{QNC} holds.  Set $a_i=-1/2+s_i/B$ for $i=1,\ldots,q$ and  $a_i\gg-1/2$ for $i=q+1,\ldots,r$.
Following the notation of \S 3, let
$\boldsymbol{s}:=\{s_1,\ldots,s_q\}=\{\sigma_i^{\mu_i},\ldots,\sigma_p^{\mu_q}\}=:\boldsymbol{\sigma}^{\boldsymbol{\mu}}$
 with $q=|\boldsymbol{\mu}|$.  Moreover, let $\mathcal{A}'$ be a
 contour going from $\infty e^{2\mathrm{i}\pi/3}$ to $\infty
 e^{-2\mathrm{i}\pi/3}$ and not intersecting
 $\mathcal{A}_{\boldsymbol{\sigma}}$.
Then, as $N\rightarrow\infty$,
\[B\KKL\left(A+BX,A+BY\right)
=\mathcal{K}^{\mathrm{soft}}_{\boldsymbol{\mu}}(X,Y)+\order{\frac{1}{N}},\]where
\begin{align}\mathcal{K}^{\mathrm{soft}}_{\boldsymbol{\mu}}(X,Y)&
=K^{\mathrm{soft}}(X,Y)+\sum_{i=1}^q\widetilde{\mathrm{Ai}}^{(i)}(X)\mathrm{Ai}^{(i)}(Y)\label{Soft1st}\\
&=\int_{\mathcal{A}_{\boldsymbol{\sigma}}}\frac{du}{2\pi\mathrm{i}}\int_{\mathcal{A}'}\frac{dv}{2\pi\mathrm{i}}
\frac{e^{-Xu+u^3/3}e^{Yv-v^3/3}}{v-u}\prod_{i=1}^p\left(\frac{v-\sigma_i}{u-\sigma_i}\right)^{\mu_i}\label{Soft2nd}\\
&= \frac{1}{(X-Y)}\left(A(X,Y)
+\sum_{i=1}^{q}\mu_i\Ait{\boldsymbol{\mu}+\mathbf{e}_i}({X})\Ai{\boldsymbol{\mu}-\mathbf{e}_i}({Y})\right)\label{Soft3rd}
\end{align}
with
\[A(X,Y)=\Ait{\boldsymbol{\mu}}({X})\frac{d\Ai{\boldsymbol{\mu}}({Y})}{dY}-\Ai{\boldsymbol{\mu}}({Y})\frac{d\Ait{\boldsymbol{\mu}}(X)}{dX}.
\]
\end{theorem}
\begin{proof}We first consider Eq.~\eqref{Soft1st}. The previous lemma implies
\[
\widetilde{\Lambda}^{(i)}(A+BX){\Lambda}^{(i)}(A+BY)=B^{-1}\widetilde{\mathrm{Ai}}^{(i)}(X){\mathrm{Ai}}^{(i)}(Y)+\order{\frac{1}{N^{2/3}}}
\quad\mbox{when}\quad i=1,\ldots,q.  \] For $i>q$, the supposition
$a_i\gg-1/2$ is equivalent to $a_i=-1/2+s_i/B$ together with
$s_i=B\bar{s}_i$ for some positive real number $\bar{s}_i$.  Then,
Lemma \ref{LemmaSoft} and Eqs  (\ref{3.a2})--(\ref{3.a3}) imply
\begin{multline*}
\widetilde{\Lambda}^{(i+q)}(A+BX){\Lambda}^{(i+q)}(A+BY)\\=
B^{-2}\widetilde{\mathrm{Ai}}^{(q)}(X){\mathrm{Ai}}^{(q)}(Y)+\order{\frac{1}{N}}=\order{\frac{1}{N^{2/3}}}
\quad\mbox{when}\quad i=1,\ldots,r-q.  \end{multline*}Moreover, it
is well known  (see \S 2) that
\[BK_N^\alpha
(A+BX,A+BY)=K^{\mathrm{soft}}(X,Y)+\order{\frac{1}{N^{1/3}}}.\] From
Proposition \ref{propKLambda} we conclude that
\[\begin{split}B\KKL(A+BX,A+BY)&=B\bKKL(A+BX,A+BY)+\order{\frac{1}{N^{1/3}}}\\
&=B{K}^{\alpha+r}_{N-r}(A+BX,A+BY)+B\sum_{i=1}^r\widetilde{\Lambda}^{(i)}(x){\Lambda}^{(i)}(y)+\order{\frac{1}{N^{1/3}}}\\
&=K^{\mathrm{soft}}(X,Y)+\sum_{i=1}^q\widetilde{\mathrm{Ai}}^{(i)}(X){\mathrm{Ai}}^{(i)}(Y)+\order{\frac{1}{N^{1/3}}}
\end{split}
\]as desired.

We now prove Eq.~\eqref{Soft2nd}.  By using integration by parts,
it is a simple exercise to show that the soft edge kernel defined in
Eq.~\eqref{3.a2} can be written as a double integral,
\[K^{\mathrm{soft}}(X,Y)=\int_{\mathcal{A}}\frac{du}{2\pi\mathrm{i}}\int_{\mathcal{A}'}\frac{dv}{2\pi\mathrm{i}}
\frac{e^{-Xu+u^3/3}e^{Yv-v^3/3}}{v-u}.\]Also, the integral
representation of the incomplete multiple Airy functions, given in
Eqs \eqref{3.a3}--\eqref{3.a4}, leads to
\[\widetilde{\mathrm{Ai}}^{(i)}(X)\mathrm{Ai}^{(i)}(Y)=\int_{\mathcal{A}_{\{s_1,\ldots,s_i\}}}\frac{du}{2\pi\mathrm{i}}\int_{\mathcal{A}'}\frac{dv}{2\pi\mathrm{i}}
{e^{-Xu+u^3/3}e^{Yv-v^3/3}}\frac{\prod_{k=1}^{i-1}(v-s_k)}{\prod_{k=1}^{i}(u-s_k)}.\]
Cauchy's theorem allows us to deform the contour
$\mathcal{A}_{\{s_1,\ldots,s_i\}}$ into
$\mathcal{A}_{\{s_1,\ldots,s_r\}}=\mathcal{A}_{\{\sigma_1,\ldots,\sigma_d\}}$.
 Eq.~\eqref{Soft1st} then gives
 \[\mathcal{K}^{\mathrm{soft}}_{\boldsymbol{\mu}}(X,Y)=\int_{\mathcal{A}_{\boldsymbol{\sigma}}}\frac{du}{2\pi\mathrm{i}}\int_{\mathcal{A}'}\frac{dv}{2\pi\mathrm{i}}
{e^{-Xu+u^3/3}e^{Yv-v^3/3}}\left(\frac{1}{v-u}+\sum_{i=1}^q\frac{\prod_{k=1}^{i-1}(v-s_k)}{\prod_{k=1}^{i}(u-s_k)}\right).\]
Eq.~\eqref{Soft2nd} is then obtained by exploiting the decomposition
\eqref{eqProdSum}.

We finally get Eq.~\eqref{Soft3rd} by multiplying
Eq.~\eqref{Soft2nd} by $(X-Y)/(X-Y)$, which gives
\[\mathcal{K}^{\mathrm{soft}}_{\boldsymbol{\mu}}(X,Y)=\frac{1}{X-Y}\int_{\mathcal{A}_{\boldsymbol{\sigma}}}\frac{du}{2\pi\mathrm{i}}\int_{\mathcal{A}'}\frac{dv}{2\pi\mathrm{i}}
\frac{e^{u^3/3-v^3/3}}{v-u}\prod_{i=1}^p\left(\frac{v-\sigma_i}{u-\sigma_i}\right)^{\mu_i}\left(-\frac{\partial}{\partial
u}-\frac{\partial}{\partial v}\right)\left(e^{-Xu+Yv}\right),\] by
integrating by parts, and by comparing the result with the
definitions of the multiple Airy functions given in Eqs
\eqref{AiryI}-\eqref{AiryII}.
\end{proof}

The ease at which the limiting kernel \eqref{Soft1st} can be deduced
from Eq.~\eqref{5.4'} contrasts with the difficulty encountered in
\cite{Baik2004} in computing the same limiting kernel from a double
contour integral representation equivalent to \eqref{QNLK}.  The
reason is that in the double contour the saddle point is such that
$w=z$ in the integrand, but then the denominator vanishes.  By
having reduced the double contour to a sum of products of single
integrals as in Eq.~\eqref{5.4'}, the complication is avoided.

We finish this subsection by showing that the kernel of the
perturbed Hermite ensemble can also be mapped to the soft edge
kernel $\mathcal{K}^{\mathrm{soft}}_{\boldsymbol{\mu}}$. By doing
this, we generalize a result previously obtained by P\'ech\'e
\cite{Peche} (i.e., Eq.~\eqref{Soft1st} for $s_1=\ldots=s_r=0$).

\begin{proposition}\label{SoftHermite}
Let $\KKH(x,y)$ be the kernel given in Proposition
\ref{PropHpertubed}.  Let also $A=\sqrt{2N}$ and
$B=1/\sqrt{2N^{1/3}}$.  Suppose $a_k=\sqrt{2N}(-1+s_k/N^{1/3})$ and
$s_k>0$ for $k=1,\ldots,q$. Suppose moreover $a_k\gg-A$ for
$k=q+1,\ldots,r$. Then, as $N\rightarrow\infty$, \[
B\KKH(A+BX,A+BY)=K^{\mathrm{soft}}(X,Y)+\sum_{i=1}^q\widetilde{\mathrm{Ai}}^{(i)}(X)\mathrm{Ai}^{(i)}(Y)+\order{N^{-1/3}}.\]
\end{proposition}
\begin{proof}
This formula is a straightforward consequence of Proposition
\ref{PropHpertubed} and the following asymptotic relations,
\[\begin{split}
\widetilde{\Gamma}^{(i)}(A+BX)&=(-1)^{N+r+1}\frac{N^{(i-1)/3}e^{N/2}}{A^{N+i-r-1}}\widetilde{\mathrm{Ai}}^{(i)}(X)\left(1+\order{N^{-1/3}}\right),\\
{\Gamma}^{(i)}(A+BX)&=(-1)^{N+r+1}\frac{A^{N+i-r}e^{-N/2}}{N^{i/3}}{\mathrm{Ai}}^{(i)}(X)\left(1+\order{N^{-1/3}}\right).\end{split}
\]
These equations can be obtained from the integral representation of
the incomplete Hermite functions by using the steepest descent
method (see for instance Lemma \ref{LemmaSoft}). We skip the detail.
\end{proof}

\subsection{From hard edge to soft edge}

It is known (see for instance \cite[\S4]{ForresterBook}) that we can
go from the hard edge to the soft by
 rescalling the spectral variables and taking the limit
 $\alpha\rightarrow\infty$:
\begin{equation}\label{hardtosoft} 2\alpha(\alpha/2)^{1/3} {K}^{\mathrm{hard}}(X,Y)=
{K}^{\mathrm{soft}}(\xi,\eta)+\order{\alpha^{-1/3}}
\end{equation}
 if
$X=\alpha^2-(2\alpha^2)^{2/3}\xi$ and
$Y=\alpha^2-(2\alpha^2)^{2/3}\eta$. This formula is a consequence of
the asymptotic relation between the Bessel and the Airy functions
\cite{Olver},
\[\left(\frac{\alpha}{2}\right)^{1/3}J_\alpha(\sqrt{X})=\mathrm{Ai}(\xi) +\order{\alpha^{-1/3}}.\]

In the subsection, we show that the mapping between the hard and the
soft edges is preserved in the perturbed Laguerre ensemble.

\begin{lemma}\label{LemmaHardtosoft}Let $X=\alpha^2-(2\alpha^2)^{2/3}\xi$ and $\alpha
\nu_i=1/2-\sigma_i/(4\alpha)^{1/3}$ with $\sigma_i>0$ for
$i=1,\ldots,d$.
 Then, as
$\alpha\rightarrow\infty$,
\[\begin{split}
\left(\frac{\alpha}{2}\right)^{(1-|\boldsymbol{\mu}|)/3}\widetilde{\mathcal{J}}^{\alpha}_{\boldsymbol{\mu}}(\sqrt{X}|\boldsymbol{\nu})
&=(-1)^{|\boldsymbol{\mu}|}\,\Ait{\boldsymbol{\mu}}(\xi|\boldsymbol{\sigma})+\order{\frac{1}{\alpha^{1/3}}}\\
\left(\frac{\alpha}{2}\right)^{(1+|\boldsymbol{\mu}|)/3}\mathcal{J}^{\alpha}_{\boldsymbol{\mu}}(\sqrt{X}|\boldsymbol{\nu})
&=(-1)^{|\boldsymbol{\mu}|}\,\Ai{\boldsymbol{\mu}}(\xi|\boldsymbol{\sigma})+\order{\frac{1}{\alpha^{1/3}}}.
\end{split}\]
\end{lemma}
\begin{proof}
The proof relies on saddle point method for both formulas.  Here we
only prove the first asymptotic development.  We have from
Eq.~\eqref{BesselI}
\[\widetilde{\mathcal{J}}^{\alpha}_{\boldsymbol{\mu}}(\sqrt{X}|\boldsymbol{\nu})
=(2\alpha)^\alpha
(1-(\alpha/2)^{1/3}\xi/\alpha)^{\alpha}\int_{\mathcal{C}_{\{0\}\cup\boldsymbol{\nu}}}\frac{dz}{2\pi\mathrm{i}}\frac{z^{|\boldsymbol{\mu}|+\alpha-1}e^{-\alpha^2z+(2\alpha^2)^{2/3}\xi
z+1/(4z)}}{\prod_{i=1}^d(z-\nu_i)^{\mu_i}}
\]
By making the change $z\mapsto z/\alpha$ and by considering
$(1-(\alpha/2)^{1/3}\xi/\alpha)^{\alpha}=e^{-(\alpha/2)^{1/3}\xi}+\order{{\alpha^{-1/3}}}$,
we get
\[\widetilde{\mathcal{J}}^{\alpha}_{\boldsymbol{\mu}}(\sqrt{X}|\boldsymbol{\nu})
=2^\alpha
e^{-(\alpha/2)^{1/3}\xi}\int_{\mathcal{C}_{\{0,\alpha\nu_1,\ldots,\alpha\nu_d\}}}\frac{dz}{2\pi\mathrm{i}}e^{\alpha
f(z)}g(z)\left(1+\order{{\alpha^{-1/3}}}\right),
\]
where
\[f(z)=-z+1/4z+\ln z\quad\text{and}\quad g(z)=\frac{e^{(4\alpha)^{1/3}\xi z}z^{|\boldsymbol{\mu}|-1}}{\prod_{i=1}^d(z-\alpha\nu_i)^{\mu_i}}.\]
The function $f(z)$ has a second order saddle point at $z_0=1/2$. We
now suppose that $\alpha \nu_i<1/2$ for all $i$. Following the
method used in proving Proposition \ref{LemmaSoft}, we deform the
closed contour $\mathcal{C}_{\{0,\alpha\nu_1,\ldots,\alpha\nu_d\}}$
in such a way that it reaches the steepest descent contour, which is
a path approaching $z_0$ from $\Im z<0$ with an angle of $-2\pi/3$
and leaving $z_0$ for $\Im z>0$ with an angle of $2\pi/3$.  Setting
\[z=-v/(4\alpha)^{1/3}+1/2\quad\text{and}\quad\alpha
\nu_i=1/2-\sigma_i/(4\alpha)^{1/3},\]we find that
\[\begin{split}
\alpha f(z)&=-2\alpha \ln 2+v^3/3+\order{{\alpha^{-1/3}}}\\
g(z)&=\frac{(-1)^{|\boldsymbol{\mu}|}2\alpha^{|\boldsymbol{\mu}|/3}}{2^{|\boldsymbol{\mu}|/3}}\frac{e^{-\xi
v}e^{(\alpha/2)^{1/3}\xi}}{\prod_{i=1}^d(v-\sigma_i)^{\mu_i}}\left(1+\order{{\alpha^{-1/3}}}\right).\end{split}\]
Returning to the integral representation of the type I multiple
Bessel function, we get
\[\widetilde{\mathcal{J}}^{\alpha}_{\boldsymbol{\mu}}(\sqrt{X}|\boldsymbol{\nu})=(-1)^{|\boldsymbol{\mu}|}\left(\frac{\alpha}{2}\right)^{(|\boldsymbol{\mu}|-1)/3}
\int_{\mathcal{A}_{\boldsymbol{\sigma}}}\frac{dv}{2\pi\mathrm{i}}\frac{e^{-\xi
v+v^3/3}}{\prod_{k=1}^d(v-\sigma_k)^{\mu_k}}\left(1+\order{{\alpha^{-1/3}}}\right)\]
Comparison with Eq.~\eqref{AiryI} shows that this is the expected
result.
\end{proof}

The above result can be applied to the incomplete Bessel and Airy
functions.  For instance, one can show  from Eqs
\eqref{BesselIncomplete} and \eqref{AiryIncomplete} that
\[ 2\alpha(\alpha/2)^{1/3}\left(\frac{X}{Y}\right)^{\alpha/2}\, \widetilde{J}^{(i)}(X)J^{(i)}(Y)=\widetilde{\mathrm{Ai}}^{(i)}(\xi)\mathrm{Ai}^{(i)}(\eta)+\order{\frac{1}{\alpha^{1/3}}}.\]
When substituted into Theorem \ref{TheoHard}, this formula together
with Theorem \ref{TheoSoft} imply the following asymptotics.

\begin{proposition}\label{PropHardtosoft}Let $X=\alpha^2-(2\alpha^2)^{2/3}\xi$,
$Y=\alpha^2-(2\alpha^2)^{2/3}\eta$ and $\alpha
\nu_i=1/2-\sigma_i/(4\alpha)^{1/3}$ for $\sigma_i>0$. Then, as
$\alpha\rightarrow\infty$,
\[2\alpha(\alpha/2)^{1/3}\mathcal{K}_{\boldsymbol{\mu}}^{\mathrm{hard}}(X,Y|\boldsymbol{\nu})
=\mathcal{K}^{\mathrm{soft}}_{\boldsymbol{\mu}}(\xi,\eta|\boldsymbol{\sigma})+\order{\alpha^{-1/3}}.\]
\end{proposition}

\section{Concluding remarks}
\subsection{The case $r=1$ at the soft edge}
Noting from (\ref{3.a3}) that $\widetilde{\mathrm{Ai}}^{(i)}(x) \to
0$ as $x \to - \infty$, we see from (\ref{3.9'}) that
$$
\widetilde{{\rm Ai}}^{(1)}(X) = \int_{-\infty}^X e^{s_1(x - X)} {\rm
Ai}(x) \, dx.
$$
Substituting in (\ref{3.a}) shows that for $r=1$, $s_1 = d$,
${\mathcal K}^{\rm soft}(X,Y)$ can be written explicitly in terms of
Airy functions
\begin{equation}\label{at2}
{\mathcal K}^{\rm soft}(X,Y) = K^{\rm soft}(X,Y) + {\rm Ai}(Y)
\int_{-\infty}^X e^{d(x - X)} {\rm Ai}(x) \, dx.
\end{equation}

In fact (\ref{at2}) has previously been derived as the soft edge
scaled correlation kernel for random matrices closely related to
(\ref{18.2}) \cite{FR1,FR2}. These are $(N+1) \times N$ complex
Gaussian matrices $\mathbf X$ with the entries of the first $N$ rows
independent complex Gaussians with mean 0 and variance 1/2, while
the entries of the final row are distributed according to this same
complex Gaussian multiplied by $\sqrt{c}$. The quantities studied
were the eigenvalues of $\mathbf X^\dagger \mathbf X$. With $c = 1/b$, such random
matrices differ from those specified by (\ref{18.2}), and the
paragraph including (\ref{3.0}) with $r=1$, only in that the final
row of $\mathbf X$ has variance $1/2\sqrt{b}$ instead of the final
column.

Results of \cite[Prop.~14]{FR1} give that with the largest $n$
eigenvalues $\{x_j\}_{j=1,\dots,n}$ of $\mathbf X^\dagger \mathbf X$
scaled by
\begin{equation}\label{s1a}
x_j \mapsto 4N + 2(2N)^{1/3} X_j
\end{equation}
and with
\begin{equation}\label{s1b}
b = \frac{1 + A }{ 2}, \qquad A = \frac{d }{ (2N)^{1/3} }
\end{equation}
the corresponding $n$-point correlation function is given by
\begin{equation}\label{at2'}
\det [ \tilde{\mathcal K}^{\rm soft}(X_j,X_k) ]_{j,k=1,\dots,n}
\end{equation}
where
\begin{equation}\label{at3}
\tilde{\mathcal K}^{\rm soft}(X,Y) = - e^{d (X-Y)} \frac{\partial }{
\partial X}
 \Big ( e^{ -d  X} \int_{-\infty}^Y e^{d v} K^{\rm soft}(X,v) \, dv \Big ).
\end{equation}

The scalings (\ref{s1a}), (\ref{s1b}) are identical to (\ref{3.0}).
Since we have shown that the soft edge scaling for the perturbed Laguerre
ensemble is independent of the parameter $\alpha$, one might suspect that it is
independent of variance being altered along a row instead of down a column.
Indeed use of the integral form of $K^{\rm soft}$ (\ref{3.a2}),
$$
 K^{\rm soft}(X,Y) = \int_0^\infty {\rm Ai}(X+t) {\rm Ai}(Y+t) \, dt,
$$
and use of integration by parts in (\ref{at3}) reproduces
(\ref{at2}), but with $X \leftrightarrow Y$, an operation which
leaves unaltered (\ref{at2'}).

\subsection{An open problem}
For the same class of random matrices which gave rise to
(\ref{at3}), the $n$-point correlation function for the smallest $n$
eigenvalues $\{x_j\}_{j=1,\dots,n}$, scaled by
$$
x_j \mapsto X_j/4N, \qquad b = 2Nw
$$
was computed in \cite[Prop.~14]{FR1}  as being equal to
$$
\det [ \tilde{\mathcal K}^{\rm hard}(X_j,X_k) ]_{j,k=1,\dots,n}
$$
where
\begin{equation}\label{su.1}
 \tilde{\mathcal K}^{\rm hard}(X,Y) = - \frac{\partial }{ \partial X}
\Big ( e^{- w X/2} \int_0^Y e^{w v/2} K^{\rm hard}(X,v) \Big
|_{\alpha = 0} dv \Big ).
\end{equation}
We have not been able to express (\ref{su.1}) in the form of
(\ref{1.18}) with $r=1$. Thus unlike the situation at the soft edge,
the eigenvalues at the hard edge distinguish the variance of the
final row of $\mathbf X$ being $1 / \sqrt{b}$ rather than the variance of
the final column.

This is not surprising when one recalls that for matrices
$
\mathbf W^\dagger \tilde{\mathbf B} \mathbf W
$
with $\mathbf W$ as in (\ref{18.2}) and $\tilde{\mathbf B}$ an $n
\times n$ positive definite matrix, the eigenvalues p.d.f.~is given
by  \cite{Chiani,SimonMM}
\begin{equation}\label{7.7}
(-1)^{N(N-1)/2} \frac{\prod_{i=1}^n b_i^N }{ \prod_{j=1}^N j! }
\frac{\prod_{j<k}^N (\lambda_j - \lambda_k) }{ \prod_{j<k}^n (b_j -
b_k) } \det \Big [ [b_j^{k-1}]_{\substack{j=1,\dots,n \\
k=1,\dots,N-n}} \quad
 [e^{-b_j \lambda_k} ]_{\substack{j=1,\dots,n \\ k=1,\dots,N}} \Big ]
\end{equation}
and is thus different in the neighbourhood of the smallest eigenvalues
to (\ref{S3}). A challenge for future studies
is to compute the $n$-point correlation for the hard edge scaling of
the smallest eigenvalues specified by (\ref{7.7}) with $\alpha = n - N \in \mathbb Z^+$
general.

\begin{acknow}
The work of P.J.F. has been supported by the Australian Research Council.
P.D. is grateful to the Natural Sciences and Engineering Research Council
of Canada for a postdoctoral  fellowship.
\end{acknow}

\end{document}